\documentclass[11pt,onecolumn]{article}
\usepackage{ISG_style}
\usepackage{times}
\usepackage{graphicx,subfigure}
\usepackage{color}
\usepackage{amsmath,amssymb} 
\usepackage{dsfont}
\usepackage{cite}
\usepackage{hyperref}
\usepackage{epstopdf}
\usepackage{url}
\usepackage{flushend}

\usepackage{graphicx}
\usepackage{multirow}
\usepackage{enumitem}
\usepackage[table]{xcolor}

\usepackage{algorithm}
\usepackage{makecell}
\usepackage{algcompatible}
\usepackage{algpseudocode}
\usepackage{mathtools}
\usepackage{verbatim}
\usepackage[font=small,justification=centering]{caption}
\ifpdf
    \graphicspath{{figures/PNG/}{figures/PDF/}{figures/}}
\else
    \graphicspath{{figures/EPS/}{figures/}}
\fi

\definecolor{mygreen}{rgb}{0.10,0.50,0.10}

\DeclareMathOperator*{\argmax}{arg\,max}  

\newcommand{\isend}{{\sf end}}
\newcommand{\buff}{{\sf Buffer}}
\newcommand{\ptdf}{{\sf PTDF}}
\newcommand{\lodf}{{\sf LODF}}
\newcommand{\trainSteps}{{\sf trainSteps}}

\newcommand{\maxAction}{{\sf maxAction}}
\newcommand{\maxReward}{{\sf maxReward}}

\newcommand{\topFiveActions}{{\sf topFiveActions}}
\newcommand{\noop}{{\sf Do\text{-}Nothing}}
\newcommand{\reco}{{\sf Re\text{-}Connection}}

\title{\bf \huge Blackout Mitigation via Physics-guided RL}
	
\date{}
	
\author{Anmol~Dwivedi \qquad   Santiago Paternain \qquad  Ali~Tajer 
\thanks{Electrical, Computer, and Systems Engineering Department, Rensselaer Polytechnic Institute, Troy, NY 12180.} 
\thanks{Our code and data are open-sourced and available on \href{https://github.com/anmold-07/Physics-Guided-Blackout-Mitigation}{GitHub.}}
} 

\begin{document}

\maketitle
\allowdisplaybreaks

\begin{abstract}
This paper considers the sequential design of remedial control actions in response to system anomalies for the ultimate objective of preventing blackouts. A physics-guided reinforcement learning (RL) framework is designed to identify effective sequences of real-time remedial look-ahead decisions accounting for the long-term impact on the system's stability.  The paper considers a space of control actions that involve both discrete-valued transmission line-switching decisions (line reconnections and removals) and continuous-valued generator adjustments. To identify an effective blackout mitigation policy, a physics-guided approach is designed that uses power-flow sensitivity factors associated with the power transmission network to guide the RL exploration during agent training. Comprehensive empirical evaluations using the open-source Grid2Op platform demonstrate the notable advantages of incorporating physical signals into RL decisions, establishing the gains of the proposed physics-guided approach compared to its black-box counterparts.
One important observation is that strategically~\emph{removing} transmission lines, in conjunction with multiple real-time generator adjustments, often renders effective long-term decisions that are likely to prevent or delay blackouts. 
\end{abstract}

\vspace{-0.1in}
\section{Introduction}
\label{sec:Introduction}

Widespread power system failures are partly due to congestion led by excessive electricity load demands.
%Specifically, under congestion, transmission line flows tend to approach the capacity limits.
%These situations compel the operation of the system in precarious proximity to its capacity limits, especially concerning transmission capacity. 
Increased and sustained congestion gradually stresses the system over time, rendering anomalies (e.g., failures) that can potentially escalate~\cite{NAPS:2004} if left unchecked. In such circumstances, system operators take real-time remedial actions to re-stabilize the system. Widely-used examples of such actions include adjusting generation ~\cite{Venkat:2008, almassalkhi:1-2014, almassalkhi:2-2014, Huang:2020} and altering network's topology~\cite{Fisher:2008, Khodaei:2010, Fuller:2012, Dehghanian:2015}. Choosing these actions involves balancing two opposing action impacts. On the one hand, an action can ensure safeguards for specific system components (e.g., adjusting flow in a line to prevent overflow). Such actions have \emph{quick} impacts but target only specific components' stability without regard for their inadvertent consequences. On the other hand, an action can ensure overall network stability. Such \emph{look-ahead}  actions have delayed impact but ensure higher robustness. Therefore, to balance these opposing impacts, forming~\emph{quick} remedial control decisions with a~\emph{look-ahead} approach is crucial for ensuring a secure and reliable power system operation.

\vspace{.05 in}
\noindent{\bf Reinforcement Learning Framework:} In this paper, we formalize a~\emph{physics-guided} reinforcement learning (RL) framework to form real-time decisions to alleviate transmission line overloads over long horizons. We focus on both \emph{discrete}-valued transmission line-switching decisions (i.e., line reconnections and removals) and \emph{continuous}-valued generator adjustments as the control decisions of interest. In this framework, finding an optimal sequence of decisions faces three key challenges. First, the cardinality of the decision space grows exponentially with the number of components and control horizon. Second, the decisions are temporally dependent, as system operational constraints limit the frequency of component interventions to maintain system security. Finally, we have a hybrid of continuous and discrete actions, quantifying the impacts of which have distinct natures.

\vspace{.05 in}
\noindent{\bf RL Methodology:}
Our RL approach is guided by 
%the physical signatures of power systems, leveraging 
two power-flow sensitivity metrics: the line outage distribution factor and the power transfer distribution factor~\cite{wood:2013}. Such physics-guided decisions facilitate more structured exploration during training RL agents and effective exploration of diverse topological configurations and generation adjustment decisions.
%enabling the discovery of effective decision combinations.
%By prioritizing topological decisions based on sensitivity factors, this structured exploration approach when combined with random generator adjustments decisions, enables the exploration of topological configurations, fostering  
Our extensive evaluations on Grid2Op~\cite{donnot:2020} -- a platform specialized for implementing RL in power transmission systems -- confirm that 
%the decisions formed by our 
physics-guided RL decisions significantly outperform the counterpart black-box RL exploration strategies.
Before specifying our approach, we review the existing literature relevant to this paper's scope. Specifically, there exist two overarching approaches for devising real-time remedial decisions: model-based and data-driven, which we review next.

\vspace{.05 in}
\noindent{\bf Model-based Approaches:} Model-based techniques such as model predictive control (MPC) focus on analytically approximating the system model and formulating a multi-horizon optimization objective to predict future states and devise control decisions. For instance, MPC approaches to optimally coordinate load shedding, capacitor switching, and transformer tap changes for long-term voltage stability are studied in~\cite{larsson:2002, Zima:2003, Hiskens:2005}, and the studies in~\cite{Carneiro:2010, almassalkhi:1-2014, almassalkhi:2-2014} alleviate line overloads by optimally coordinating generation adjustment and energy storage schedules. While model-based approaches offer increased control accuracy by ensuring adherence to system constraints, they require access to an accurate system model, which can be difficult for highly complex systems. Furthermore, characterizing optimal policies becomes particularly challenging when coordinating decisions of different natures, such as discrete line-switching and continuous generation adjustments. Such hybrid decision spaces pose analytical challenges to optimizing decisions within a unified optimization objective.

\vspace{.05 in}
\noindent{\bf Data-driven Approaches:} In another direction, there are data-driven techniques, such as deep RL. In these approaches, the objective is to learn decision policies directly via sequentially interacting with the system model. 
%These interactions learn a model to address the challenge of analytically characterizing the system. 
{For example, model-free RL algorithms have been widely employed to efficiently address various challenges in power systems, including system stability control \cite{Ernst:2004}, vulnerability analysis \cite{Yan:2017}, voltage control \cite{Duan:2020}, and risky fault-chain identification \cite{GRNN:2024}, among others.} Furthermore, the flexibility to model decisions via a Markov decision process (MDP) facilitates accommodating diverse remedial control types and serving deep RL as a promising candidate for overload management~\cite{kelly:2020, marot:2020, marot:2021}.

\noindent{\bf Deep RL Approaches:} There exist several deep RL approaches designed for
%to devise real-time remedial controls in the context of 
transmission line overload management. These approaches can be categorized based on the nature of their decisions. For instance, the study in~\cite{Lan:2020} focuses on~{a subset of bus-split} topological actions and proposes a guided RL exploration training approach that selects control actions guided by the highest $Q$-values recommended by a dueling $Q$-network~\cite{wang2016dueling}. %via extensive simulations. 
{To accommodate the exponentially many bus-split topological actions, the study in~\cite{ICLR:2021} employs graph neural networks combined with hierarchical RL~\cite{sutton:1999}. This approach learns a target network topology based on the current MDP state and leverages bus-splitting to achieve the predicted objective.} Alternatively, the approach of~\cite{AAAI:2023} pre-selects~\emph{effective} topological actions via extensive power-grid simulations and subsequently employs policy gradient-based RL algorithms using the pre-selected action space to find a decision policy. {The study in~\cite{Matavalam:2023} proposes a curriculum-based approach for training RL agents, emphasizing the significance of integrating domain knowledge into RL frameworks. Building upon this work,~\cite{meppelink:2023} proposes a hybrid approach that combines the curriculum-based and Monte-Carlo tree search (MCTS)-based~\cite{silver:2016} approaches to enhance exploring secure remedial control actions.} 

\vspace{.05 in}
\noindent{{\bf Our contributions:} The common aspect of all the existing RL approaches discussed earlier is that (i)~they focus entirely on bus-splitting actions, and (ii) they do not incorporate any physical power system signals to guide exploration decisions. The contributions of this paper can be summarized as follows.
\begin{itemize}[leftmargin=8pt]
    \item {\bf Physics-guided:} Our approach is physics-guided and leverages~\emph{sensitivity factors} to structure exploration during training. The experiments establish the gains of explicitly integrating network topology information into RL. This also alludes to the potential for broader applicability to other topological control actions, such as bus-splitting.
    \item {\bf Active line-switching:} Active line switching and particularly line~\emph{removal} actions, have been unexploited in designing action spaces. This is perhaps primarily due to concerns that line removal may decrease power transfer capabilities, thereby increasing the risk of cascades. However, we demonstrate that while line removals may momentarily decrease transfer capabilities,~\emph{strategically} removing lines over a planned horizon can prevent or delay cascades, proving to be an effective remedial action.
    \item {\bf Hybrid action space:} Our framework~\emph{jointly} optimizes continuous generator dispatch and discrete line switching within a single RL agent. This integration, despite its modeling and optimization challenges due to mixed action spaces, demonstrably improves power grid resilience by drastically increasing the system's survival time. We note that adopting a hybrid action space is also investigated in ~\cite{dorfer:2022}. Nevertheless, its distinction is that it decouples the decisions, designing the discrete and continuous (generation adjustment) independently.
\end{itemize}}

\vspace{-0.1in}
\section{Problem Formulation}
\label{sec: Problem Formulation}

\subsection{System Model}
\label{sec: System Model}
%\SP{The notation $(A_\ell[n])$ is because of the changing topology of the network? Is $A$ the adjacency matrix? Is ``denoted via'' correct English? Isn't it more common to use ``denoted as'' or ``denoted by''?}
Consider a power transmission system consisting of $N$ buses indexed by $[N]\triangleq \{1,\dots, N\}$ and $L$ transmission lines indexed by $[L]\triangleq \{1,\dots, L\}$. 
We denote the power and current flows in line $\ell \in [L]$ at the discrete-time instant $n\in\N$ by $F_{\ell}[n]$ and $A_{\ell}[n]$, respectively. Accordingly, we define 
%in all transmission lines $\ell \in [L]$ can be compactly denoted by vectors via $\bF[n]\;(\bA[n]) \in \R^{L}$
\begin{equation*}
\begin{aligned}
    \label{eq: flows}
        \bF[n] \dff [F_{1}[n], \dots, F_{L}[n]]^\top,\ 
        \bA[n] \dff [A_{1}[n], \dots, A_{L}[n]]^\top\ .
\end{aligned}
\end{equation*} 
Each line $\ell \in [L]$ is constrained with maximum power and current flows, denoted by $F_{\ell}^{\sf max}$ and $A_{\ell}^{\sf max}$, respectively.
The transmission system has $G$ dispatchable generators serving $D$ loads. The generators output are injected in buses $\mcG\subseteq [N]$ indexed by $\mcG\triangleq \{g_j: j\in[G]\}$ and the loads draw power from buses $\mcD\subseteq [N]$ indexed by $\mcD\triangleq \{d_k:k \in [D]\}$. We denote the real-power injected by generator $g_j \in [G]$ by $G_{j}[n]$ (in MWs) and the real-power demand by load $d_{k} \in [D]$ (in MWs) by $D_{k}[n]$. For all other buses $i\in[N]$ not connected to a generator or a load component we have $G_i[n]=0$ and $D_i[n]=0$. Accordingly, we define generator and load vectors
\begin{equation*}
%\small
\begin{aligned}
    \label{eq: injections}
        \bG[n] \dff [G_{1}[n], \dots, G_{N}[n]]^\top,\ 
        \bD[n] \dff [D_{1}[n], \dots, D_{N}[n]]^\top .
\end{aligned}
\end{equation*} 
We refer to the net-power injection in bus $i \in [N]$ by $\bP[n] \dff \bG[n] - \bD[n]$. We assume that each line $\ell \in [L]$ is specified by $F$ features (e.g., flows and status) and each bus $i\in[N]$ is specified by $H$ features (e.g., demand and generation). For compact representation, we concatenate all these features and represent them by a system state vector, which at time $n$ is denoted by $\bX[n]\in\R^{L\cdot N+F\cdot H}$. Finally, we define the~\emph{operational} status of each line $\ell\in [L]$ as a binary variable $\Omega_{\ell}[n]$ such that $\Omega_{\ell}[n]=1$ if line $\ell$ is operational at time $n$, and otherwise $\Omega_{\ell}[n]=0$.
% \begin{equation}
% \small
% \begin{aligned}
%     \label{eq: status}
%     \Omega_{\ell}[n] \dff \begin{cases}
%         1 & \text{if line $\ell$ is operational at time $n$} \\
%         0 & \text{if line $\ell$ is not operational at time $n$} 
%     \end{cases}\ .
% \end{aligned}
% \end{equation} 
We denote the set of lines operational at $n$ by $\mcL[n] \dff \{\ell \in [L]\;:\;\Omega_{\ell}[n] = 1\}$.
% \begin{align}
%     \label{eq: operational set}
%     \mcL[n] \dff \{\ell \in [L]\;:\;\Omega_{\ell}[n] = 1\}\ .
% \end{align}

%\vspace{-0.15in}
\subsection{Blackout Mitigation Model}
\label{sec: Cascading Failure Model}
Transmission systems are constantly prone to being stressed by adverse internal (e.g., line failures) and external (e.g., excessive demand) conditions. Without proper remedial actions, sustained stress can lead to failures, further compounding the stress and resulting in more failures. Prolonged periods of such cascading failures can lead to circumstances in which the system cannot meet the load demand. We refer to such circumstances as~\emph{blackouts}. Minimizing the likelihood of such blackouts necessitates devising actions that can swiftly react to emerging system stress to confine its impact.

To mitigate the adverse consequences of cascading failures in the network, our objective is to maximize the system's survival time (ST). The ST of the system %of a horizon of $T$ 
is defined as the first instance at which a system blackout occurs\cite{donnot:2020, Lan:2020, ICLR:2021}. 
%\ATAD{refs?} 
Maximizing the ST is facilitated by the following two key sets of actions that aim to reduce stress via controlling line flows.
\subsubsection{Line-Switching Actions} 
\label{sec: Line-Switch Decisions}
Line-switching decisions pertain to removing or reconnecting lines to alleviate overloads. For each line $\ell$ at a given time $n$, we define the binary decision variable $W_{\ell}[n] \in \{0, 1\}$ such that $W_{\ell}[n]=0$ indicates removing line $\ell$ while $W_{\ell}[n]=1$ indicates reconnecting it. Accordingly, we define the decision vector at time $n$ by 
%set $\mcC_{\sf line}[n]$ defined via
\begin{align}
    \label{eq: line decisions}
  %  \mcC_{\sf line}[n] \dff \{W_{\ell}[n]:\ell \in [L]\}\ ,
   % \mcC_{\sf line}[n] \dff [ W_1[n], \dots, W_L(n)]\ .
    \bW_{\sf line}[n] \dff [ W_1[n], \dots, W_L(n)]^{\top}\ .
\end{align} 
Such line-switching decisions have operational constraints for implementing them. Specifically, we consider two constraints for line failures and line-switching decisions. First, once we apply a line-switching decision on a line, a mandated downtime period $\tau_{\rm D}$ must elapse before this line can be again considered for a line-switching decision. Secondly, when a line fails naturally (e.g., due to severe overloads), it can be reconnected only after a mandated downtime period of $\tau_{\sf F}$, where $\tau_{\rm F} \gg \tau_{\rm D}$. Furthermore, different lines have different line-switching costs. We define $c^{\sf line}_{\ell}$ as the cost of line-switching for line $\ell$. Hence, the system-wide cost of line-switching over a horizon~$T$ is
\begin{equation}
\label{eq: cost_line}
C_{\sf line}(T) \dff \sum_{n=1}^{T}  \sum_{\ell=1}^{L} c^{\sf line}_{\ell}\cdot W_{\ell}[n] \  .
\end{equation}
%Following a failure, the flows in other healthy components $\ell \in \mcL[n+1]$~\emph{may} exceed their rated line limits $F_{\ell}^{\sf max}$ rendering additional overloads within the network, potentially triggering a cascading effect. 

\subsubsection{Generator Adjustments} 
\label{sec: Generator Adjustment Decisions}

Alongside line-switching, real-time generator injection adjustments are also instrumental in alleviating overloads. We denote the incremental adjustment to the injection of generator $g_j\in\mcG$ at time $n$ by $\Delta G_{j}[n] \dff G_{j}[n+1] - G_{j}[n]$
% \begin{align}
%     \label{eq: genadjust}
%     \Delta G_{j}[n] \dff G_{j}[n+1] - G_{j}[n]\ ,
% \end{align}
and accordingly, define
\begin{align}
    \label{eq: gen decisions}
    %\mcC_{\sf gen}[n] \dff \{\Delta G_{j}[n]: g_{j}\in [G]\}\ ,
    \Delta \bG_{\sf gen}[n] \dff [\Delta G_1[n], \dots, \Delta G_N[n]]^{\top}\ .    
\end{align} 
%where $\Delta G_{j}[n] \in \R$ signifies the real-time adjustment to generator $g_{j}$'s injection at time $n$ between consecutive time instances as $\Delta G_{j}[n] \dff G_{j}[n+1] - G_{j}[n]$. 
Such adjustments are constrained by generation ramping constraints, enforcing a maximum adjustment level on each generator. We denote the maximum ramp-rate associated with generator $g_j\in\mcG$ by $\Delta G_{j}^{\sf max}$. Finally, we consider a linear generation cost such that the per-unit cost of adjusting the generation $g_j\in \mcG$ is $c^{\sf gen}_j$ (in $\$$/MW).

Therefore, the cumulative system-wide cost incurred by generation adjustments is %over a horizon of $T$ is
\begin{equation}
\label{eq: cost_gen}
    C_{\sf gen}(T) \dff \sum_{n=1}^{T} \sum_{j=1}^{G} c^{\sf gen}_{j}\cdot \Delta G_{j}[n] \ .
\end{equation}

\vspace{-0.2in}
\subsection{Maximizing Survival Time}
\label{sec: Problem Statement}
{Our objective is to constantly monitor the system and, when there is evidence of mounting stress (e.g., imminent overflows), initiate a series of flow control decisions (line switching and generation adjustment) that maximizes the survival of the system over a horizon of at least $T$.}
Such decisions are highly constrained with (i) the decision costs $C_{\rm line}(T)$ and $C_{\rm gen}(T)$, and (ii) system operational constraints of generation ramping and line-switching downtime periods $\tau_{\rm N}$ and $\tau_{\rm F}$. 
\begin{comment}
\begin{equation}
\label{eq:OBJ1}
\mcP:\left\{
%(\beta_{\rm line},\beta_{\rm gen}): \quad \quad
\begin{array}{cl}
     \displaystyle \max_{ \{\bar\bW_{\rm line}, \bar{\Delta \bG_{\rm gen}}\}}  & \suvt(T) \\
     {\rm s.t.}  & C_{\rm gen}(T) \leq \beta_{\rm gen}\\
      &            C_{\rm line}(T) \leq \beta_{\rm line}\\
      & \mbox{\footnotesize \sf Operational constraints in}~\ref{sec: Line-Switch Decisions},~\ref{sec: Generator Adjustment Decisions}
\end{array}\right. \ ,
\end{equation}
where ST$(T)$ denotes the survival time of the system over a horizon of $T$. 
\end{comment}

To quantify the system's survival time over a horizon $T$, denoted by ST$(T)$, we use a proxy that quantifies and tracks the risk margins of the transmission lines. Minimizing such risk margins minimizes the possibilities of line overloads, a key factor that contributes to maximizing ST$(T)$.
Specifically, for each line $\ell\in[L]$ at time $n$ we define the \emph{risk margin} as
\begin{align}
    \label{eq: margin}
    \rho_{\ell}[n] \dff \frac{A_{\ell}[n]}{A_{\ell}^{\sf max}}\ ,
\end{align}
based on which the transmission line $\ell$ is~\emph{overloaded} when $\rho_{\ell}[n] \geq 1$. Accordingly, we define the risk margin vector $\brho[n]\triangleq [\rho_1[n],\dots, \rho_{L}[n]]^{\top}$. Assessing the ST$(T)$ based on these risk margins is motivated by the fact that the system's susceptibility to disruptions stemming from various conditions may result in line failures. Following a failure, the line risk margins $\rho_{\ell}[n]$ associated with the healthy lines can be potentially compromised, rendering additional overloads within the system and raising the possibility of a cascading effect.  Hence, minimizing the aggregate risk margins of all lines over a horizon decreases the risk of overloading healthy lines, which in turn increases the likelihood of a higher ST.

We use risk margins to determine when the system is in a~\emph{critical} state, i.e., has reached a level that requires remedial interventions. This is specified by the rule
\begin{align}
\label{eq:riskthreshold}
    \max_{i \in [L]} \quad \rho_{i}[n] \geq \eta\ .
\end{align}
When the condition in~\eqref{eq:riskthreshold} is satisfied, subsequently, our objective is to sequentially form the decisions $\bar\bW_{\rm line}\triangleq \{\bW_{\sf line}[n]:n\in\N\}$ and ${\Delta \bar\bG_{\rm gen}}\triangleq \{\Delta \bG_{\sf gen}[n]:n\in\N\}$ such that they maximize the survival time of the system subject to the controlled decision costs and specified operational constraints. These decisions can be obtained by solving
\begin{equation}
\label{eq:OBJ2}
\mcP: \left\{
%(\beta_{\rm line},\beta_{\rm gen}): \quad \quad
\begin{array}{cl}
     \displaystyle 
     \min_{\{\bar\bW_{\rm line}, {\Delta \bar\bG_{\rm gen}\}}}
     & \displaystyle \sum_{n=1}^{T}  \sum_{\ell=1}^{L} \rho_{\ell}[n]\\
     {\rm s.t.}
     & C_{\rm gen}(T) \leq \beta_{\rm gen}\\
     & C_{\rm line}(T) \leq \beta_{\rm line}\\
     & \mbox{\footnotesize Operational constraints in}
     ~\ref{sec: Line-Switch Decisions},
     ~\ref{sec: Generator Adjustment Decisions}
\end{array}\right. \ .
\end{equation}

{In~\eqref{eq:OBJ2}, $\beta_{\rm gen}$ controls the total allowable cost for interventions on generator outputs and while $\beta_{\rm line}$ restricts the total cost associated with line-switching operations. These ensure that the actions are economically viable within the limits set by the system operator. Furthermore, the operational constraints in Section~\ref{sec: Line-Switch Decisions} ensure that the decisions to switch transmission lines are compatible with maintaining system security, and the constraints in Section~\ref{sec: Generator Adjustment Decisions} mandate that generation adjustments $\Delta \bG_{\sf gen}[n]$ adhere to predefined ramp rates.}

While solving $\mcP$ dynamically over time, whenever the system's risk is deemed small enough that remedial actions are not necessary, we terminate our line-switching and generation adjustment decisions, and the system returns to its normal operation. To formalize this, we use the rule $\max_{i \in [L]} \quad \rho_{i}[n] \leq \nu $
% \begin{align}
% \label{eq:riskthreshold2}
%     \max_{i \in [L]} \quad \rho_{i}[n] \leq \nu\ 
% \end{align}
to determine when to terminate the remedial actions. To ensure timely remedial control actions, it is common to set $\eta \in [0.9, 1.0]$ and $\nu \in [0, 0.9]$. %\AT{Add a comment comparing the two thresholding rules.}
We note that solving $\mcP$ assumes the fulfillment of demand $\bD[n]$ at all times $n$ without resorting to load-shedding as a possible control action. This is a deliberate choice that emphasizes the entire focus on the optimal design of other actions (line-switching and generation adjustment) while ensuring that the system demand is met. %Implementing control measures that lead to scenarios where the demand cannot fulfilled, owing to operational constraints, renders a system failure compromising survival time. 

\section{Blackout Mitigation as an MDP}
\label{sec: Cascading Failure Mitigation as a mdp}

%\subsection{Motivation}
%\label{sec: Sequential Control Measures}

Sequentially finding the decisions 
%$\bar\bW_{\rm line}= \{\bW_{\sf line}[n]:n\in\N\}$ and ${\Delta \bar \bG_{\rm gen}}=\{\Delta \bG_{\sf gen}[n]:n\in\N\}$ 
that maximize the ST faces two broad types of challenges.\\
{\bf 1. Temporally Dependent Decisions:} The first pertains to quantifying the impact of the current decision on future decisions. For instance, at time $n=1$, the system receives the demand $\bD[1]$ and itcan decide about control actions $\bW_{\sf line}[1]$ or $\Delta\bG_{\sf gen}[1]$, based on their anticipated impact on maximizing ST. Forming these decisions
faces two potentially opposing impacts. The first pertains to its~\emph{immediate} adjustments of risk margins $\sum_{\ell=1}^{L} \rho_{\ell}[2]$ in the context of meeting demand $\bD[2]$, while the second considers the lasting influence on all~\emph{future} risk margins in accommodating upcoming demands $\{\bD[n]\}_{n=3}^{T}$. For instance, a line-switching decision affects the future availability of the lines for switching, which are constrained by downtime operational constraints $\tau_{\rm F}$ and $\tau_{\rm D}$.\\
{\bf 2. Computational Complexity:} The complexity of identifying optimal line-switching (discrete) and generation adjustments (continuous) decisions grows exponentially with the number of lines $L$, the number of generators $G$, and the target horizon $T$. This complexity is further compounded by the need to adhere to operational constraints at all times. To address the challenges of solving~\eqref{eq:OBJ2}, we design a~\emph{sequential}, agent-based, and physics-guided RL algorithm. %that~\emph{sequentially} determines remedial control actions over time.

%\vspace{-0.15in}
\subsection{Modeling Failure Mitigation as an MDP}
\label{sec: Modeling Failure Mitigation as a MDP}

At any instance $n$, we have access to the system's states $\{\bX[m]:m\in[n]\}$, based on which we determine the line-switching and generation adjustment decisions. Implementing these remedial decisions results in partly deterministic outcomes -- reflecting the influence of the implemented decision on the system state -- and partly stochastic, representing changes arising from the randomness associated with unknown future demands. Hence, we employ a Markov decision process (MDP) to effectively model the stochastic interactions between the system's states and our remedial control actions over time. Such a model is formally characterized by the MDP tuple $(\mcS, \mcA, \P, \mcR, \gamma)$, specified next.
%\AT{Jus saying we approximate is not enough. You need to justify why this is a good approximation.} 

\subsubsection{MDP State Space $\mcS$} Based on the system's state $\bX[n]$, which captures the line and bus features, %crucial for decision making. However, the absence of information regarding future load demands $\{\bD[k]\}_{k=n+1}^{T}$ obscures evidence of escalating stress on the system. To address this, 
we denote the MDP state at time $n$ by $\bS[n]$, which is defined as a moving window of the states of length $\kappa$, i.e., 
%\vspace{-4pt}
\begin{equation}
\begin{aligned}
    \label{eq: mdp state}
    \bS[n] \dff \left[ \bX[n - (\kappa-1)], \dots, \bX[n] \right]^{\top}\ ,
\end{aligned}
%\vspace{-6pt}
\end{equation} 
%that encompasses the past $\kappa$ system states and refer to $\bS[n]$ as  
where the state space is $\mcS = \R^{\kappa\cdot(L\cdot N+F\cdot H)}$. Leveraging the temporal correlation of demands, decisions based on the MDP state $\bS[n]$ help predict future load demands.

\subsubsection{Action Space $\mcA$}
\label{sec: Action Space}

%Upon receiving a new system state $\bX[n]$, we infer the MDP state $\bS[n]$ via~\eqref{eq: mdp state} to facilitate effective remedial control actions. To formalize this process, 
We denote the action space by $\mcA \triangleq \mcA_{\sf line} \bigcup \mcA_{\sf gen}$, where $\mcA_{\sf line}$ and $\mcA_{\sf gen}$ are the spaces of line-switching and generation adjustment actions, respectively.

Action space $\mcA_{\sf line}$ includes two actions for each line $\ell \in [L]$ associated with \emph{reconnecting} and \emph{removing} it. Besides these $2L$ actions, we also include a \emph{do-nothing} action to accommodate the instances at which (i)~the mandated downtime period $\tau_{\rm D}$ makes all line-switch actions operationally infeasible; or (ii)~the system's risk is sufficiently low. This allows the model to determine the MDP state at time $n+1$ solely based on the system dynamics driven by changes in load demand $\bD[n+1]$. 
%\noindent {\bf Generator Action Set $\mcA_{\sf gen}$:} 
In action space $\mcA_{\sf gen}$, incremental generator adjustments $\Delta G_{j}[n]$ 
%in~\eqref{eq: genadjust} 
are~\emph{continuous}. To harmonize these actions with the discrete set $\mcA_{\sf line}$, we discretize generator adjustment decisions $\Delta \bG_{\rm gen}[n]$ in~\eqref{eq: gen decisions} for each $g_j \in \mcG$ when solving $\mcP$ in~\eqref{eq:OBJ2}. To ensure that the discretized outputs satisfy ramping constraints $\Delta G_{j}^{\rm max}$, we select
%we carefully alter 
$\Delta \bG_{\rm gen}[n]$ such that %specified in~\eqref{eq: gen decisions}. 
%Specifically, we introduce ${\Delta \tilde G_{j}}[n]$, representing a discretized incremental generator adjustment, that imposes constraints on the incremental adjustments to $\delta$
%\AT{should we change $\delta \leq $ to $\delta =$?}
\begin{equation}
%\small
\begin{aligned}
\label{eq: generator discrete}
\Delta \tilde G_{j} \in \{-\delta,\;0,\;+\delta\}\ , \; \;\; \delta =  \min \{ \Delta G_{j}^{\rm max}:g_j\in\mcG\}\ .
\end{aligned}
\end{equation} 
Choosing $\delta$ larger than the choice in~\eqref{eq: generator discrete} excludes the generators with ramping constraints smaller than $\delta$ from actively participating in the generator decisions. In contrast, choosing a smaller value for $\delta$ excludes larger adjustments that satisfy the ramping constraints and can expedite directing the flows in the desired trajectories. 
%Choosing $\delta$ to satisfy the ramping constraints of all generators ensures all generators are candidates for generating adjustment constraints at every instance. 
%the active participation of all generators, as the discretized output $\delta$ is chosen to be smaller than the maximum ramp-rates of all $g_j \in \mcG$.
Furthermore, to ensure that generation adjustments are operationally viable at all times $n$, we design each action $a[n] \in \mcA_{\sf gen}$ as a~\emph{feasible} combination involving adjustments to the injections of~\emph{multiple} generators.
\begin{equation}
\begin{aligned}
\label{eq: generator action}
    \sum_{j=1}^{G} \; \Delta \tilde G_{j}[n] = 0\ , \; {\rm where} \;\; \Delta \tilde G_{j}[n] \in \{-\delta,\; 0,\;+\delta\}\ .
\end{aligned}
%\vspace{-6pt}
\end{equation}

\subsubsection{Stochastic Transition Dynamics $\P$} After an action $a[n] \in \mcA$ is taken at time $n$, the  MDP state $\bS[n]$ transitions to the next state $\bS[n+1]$ according to an unknown transition probability kernel $\P$, i.e., 
\begin{equation}
%\small
\begin{aligned}
\label{eq: system dynamics}
    \bS[n+1] \sim \P(\bS \;|\;\bS[n], a[n])\ ,
\end{aligned}
\end{equation} where $\P$ captures the system dynamics influenced by both the random future load demand and the implemented action $a[n]$.

\subsubsection{Reward Dynamics $\mcR$}
% variational form, aggregate cost with competing objectives
To capture the immediate effectiveness of taking an action $a[n]\in \mcA$ in any given MDP state $\bS[n]$, we define an instant reward function
%$r[n]\in\mcR$
\begin{align}
%\begin{aligned}
\label{eq: reward}
 r[n]   \dff \sum_{\ell=1}^{L} \left( 1 - \rho_{\ell}^{2}[n]  \right) 
     - 
 \mu_{\sf gen} \left( \sum_{j=1}^{G} c^{\sf gen}_{j} \cdot \Delta \tilde G_{j}[n] \right) - \mu_{\sf line} \left( \sum_{\ell=1}^{L} c^{\sf line}_{\ell}\cdot  W_{\ell}[n]\right)\ ,
%\end{aligned}
\end{align} 
which is the decision reward associated with transitioning from MDP state $\bS[n]$ to $\bS[n+1]$, where the constants $\mu_{\sf gen}$ and $\mu_{\sf line}$ are associated with the cost constraints $\beta_{\rm gen}$ and $\beta_{\rm line}$ introduced in~\eqref{eq:OBJ2}, respectively. The inclusion of parameters $(\mu_{\sf gen}, \mu_{\sf line})$ allows us to flexibly model different cost constraints, reflecting diverse economic considerations in power systems. Greater values for the parameters $\mu_{\sf gen}$ and $\mu_{\sf line}$ in~\eqref{eq: reward} promote solutions that satisfy stricter cost requirements. To avoid tuning these parameters manually one can resort to primal-dual methods~\cite{paternain2019constrained}, where the parameters $(\mu_{\sf gen}, \mu_{\sf line})$ can be adapted based on the extent of constraint violation.

%\SP{natural question: how to select these parameters?} \AT{need a discussion about how these are chosen and how they create tradeoffs} 

For any decision-making policy $\pi$, the aggregate reward when the system is initiated at state $\bS[1]=\bS$ %taking action $a$ and thereafter, following an control action selection policy $\pi$ for $n\geq 1$ 
can be characterized by the state-action value function, given by %\AT{should we have $a[0]=a$ in the next equation?}
\begin{align}
%\footnotesize
%\begin{aligned}
\label{eq: value function}
Q_{\pi}(\bS, a)  \dff  \E\left[\sum_{n=1}^{T}\gamma^{n}r[n]\;\Big|\;\bS[1] = \bS, a[1] = a, a[n]_{n\geq2} = \pi(\bS[n]) \right]\ ,
%\end{aligned}
\end{align} 

where $\gamma\in\R_+$ is the discount factor that decides how much future rewards are favored over instant rewards.
%, and $\pi(\bS[n])$ denotes the control action taken in MDP state $\bS[n]$. 
Hence, finding an~\emph{optimal} decision policy $\pi^{*}$ can be found by solving\cite{bellman:1957} %$\mcP_{2}$
\begin{align}
    \label{eq: optimal policy}
    \mcP_{2}: \quad \pi^{*}(\bS) \dff \argmax_{\pi} \; Q_{\pi}(\bS, \pi(\bS))\ .
\end{align} 

\section{Physics-guided RL Solution}
\label{sec: Physics-guided Approach to Solving}

%\AT{some statements in this section can be more concise.}

\subsection{Motivation}
\label{sec:Motivation}

Determining an optimal policy $\pi^{*}$~requires knowing the transition dynamics, captured by $\P$, for the high-dimensional MDP state space $\bS \in \mcS$. 
When an accurate system dynamics model is available, model-based approaches (e.g., dynamic programming) can be effective for determining an optimal policy $\pi^{*}$. However, obtaining accurate models can be computationally prohibitive due to extensive offline simulator interactions, especially as the system size scales up. In such cases,~\emph{model-free} RL is effective in finding an optimal policy without explicitly learning the system model. A crucial aspect in the design of such algorithms is a comprehensive exploration of the MDP state space $\mcS$ to learn the expected decision utilities, e.g., to obtain accurate $Q$-value estimates.

For addressing these, $Q$-learning with function approximation~\cite{Mnih:2015} serves as an effective value-based~\emph{model-free off-policy} RL approach
~\cite{tsitsiklis:1997, sutton:2018}. Such methods entail dynamically updating a behavior policy $\pi$, informed by a separate exploratory policy such as $\epsilon$-greedy~\cite{sutton:2018}. 
While $Q$-learning algorithms with random $\epsilon$-greedy exploration are effective in many domains~\cite{Mnih:2015}, their application to power systems encounters a significant limitation. {Specifically, without properly selecting exploration actions, a generic exploration policy often inadvertently selects actions that quickly lead to severe overloads and, consequently, blackouts. This results in the agent interacting with the MDP for significantly fewer time steps, specifically instances when $\max_{i \in [L]} \rho_{i}[n] \geq \eta$, as the exploration process terminates upon reaching such critical MDP states, preempting a comprehensive exploration of the state space $\mcS$ (more details in Section~\ref{sec: Effect of Agent Type on Survival Time}(a)). Consequently, this limitation adversely affects the accuracy of $Q$-value predictions for unexplored states in the MDP, rendering a highly suboptimal remedial control policy design.} To address this, we design a~\emph{physics-guided} exploration policy that judiciously exploits the underlying structure of the MDP state $\mathcal{S}$ and action $\mathcal{A}$ spaces for a more accurate $Q$-value prediction.

\subsection{Action Mapping via Sensitivity Factors}
\label{sec: Action Mapping via Sensitivity Factors}

Random exploratory actions,~\emph{particularly} those involving the network topology $\mcA_{\sf line}$ over long horizons $T$, lead to~\emph{rapid} system-wide failures~\cite{ICLR:2021, Lan:2020}. This is because topological actions force an abrupt change in the system state by redistributing flows after a network topological change, compromising risk margins $\rho_{\ell}$ and exposing the system to potential cascading failures. This prevents the agent from exploring a diverse set of MDP states that are significantly different from the initial state $\bS[1]$, in turn, limiting action diversity, and subsequently, adversely impacting the learned~policy $\pi$. To address this, we leverage power-flow~\emph{sensitivity factors}~\cite{wood:2013} to guide the selection of the exploration decisions.

Sensitivity factors facilitate expressing the mapping between MDP states $\mcS$ and actions $\mcA$ derived by linearizing the system about the current operating point. Their primary function is to recursively approximate line flows based on power injection fluctuations.
%the change in flows $F_{\ell}$ for any line $\ell \in [L]$ due to variations in power injections at any bus $j \in [N]$. 
This is facilitated through linearizing power-flow equations
%\footnote{(i) The system is lossless; (ii) voltage magnitudes are constant ($1$ p.u.); and (iii) voltage phase angles between neighboring buses are small at all times.} 
to effectively assess the impact of an action $a\in\mcA$ on all line flows $F_{\ell}$ and, consequently, the reward $r\in \mcR$. Two distinct types of sensitivity factors can be characterized based on action $a\in\mcA$ under consideration.

\subsubsection{The Power Transfer Distribution Factor}
%Effect of Generator Adjustments on Flows via $\ptdf$
 The power transfer distribution factor (PTDF) matrix $\ptdf\in\R^{L \times N}$ represents the sensitivity of flows $F_{\ell}[n]$ in any line $\ell \in \mcL[n]$ to generator adjustments $\Delta \tilde G_{j}[n]$ on any bus $j$ via~\footnote{We assume that the remaining generation is made up by the reference bus.}
\begin{equation}
\begin{aligned}
    \label{eq: ptdf}
    F_{\ell}[n+1] \approx F_{\ell}[n] + \ptdf_{\ell, j}[n] \cdot \Delta \tilde G_{j}[n]\ .
\end{aligned}
\end{equation}
However, since action $a[n] \in \mcA_{\sf gen}$ involves adjustments to multiple generators~\eqref{eq: generator action}, approximation~\eqref{eq: ptdf} is less applicable. Nevertheless, by leveraging the linearity of PTDF, the influence of each action on flows can be approximated via\footnote{Note that this assumes that no generator will exceed its maximum.}
\begin{equation}
\begin{aligned}
    \label{eq: ptdf comb}
    F_{\ell}[n+1] \approx F_{\ell}[n] + \sum_{j=1}^{G}~\ptdf_{\ell, j}[n] \cdot \Delta \tilde G_{j}[n]\ .
\end{aligned}
\end{equation} %Here $\Delta G_{i} \in \{-\delta,\; 0,\;+\delta\}$.

\subsubsection{The Line Outage Distribution Factor}
%Effect of Line-Switch on Flows via $\lodf$} 
Analyzing the effect of line-switch actions, in particular line removals, can be modeled via line outage distribution factors (LODF). Matrix $\lodf\in\R^{L \times L}$, constructed~\emph{from} $\ptdf$, represent sensitivities of flows in line $\ell$ to a removal of line $k\in \mcL[n]$
\begin{align}
    \label{eq: lodf}
    F_{\ell}[n+1] \approx F_{\ell}[n] + \lodf_{\ell, k}[n] \cdot F_{k}[n]\ ,
\end{align} where $F_{k}[n]$ denotes the pre-outage flow in line $k$. The effect of line reconnections (derived from PTDF) is analyzed in~\cite{Sauer:2001}.
%Therefore, our objective is to leverage linear sensitivity factors to~\emph{efficiently} identify~\emph{effective} actions $a[n] \in \mcA$ within the design of the exploratory policy. In the context of problem $\mcP_{2}$ in~\eqref{eq: optimal policy}, actions deemed effective at time $n$ are those that decrease line margins across all lines $\sum \rho_{\ell}[n]$. 

\begin{algorithm}[t] 
    \caption{Finding Control Policy $\pi_{\btheta}$ via Deep $Q$-learning}
    \label{alg:train}
    \begin{algorithmic}[1]
        %\scriptsize
        %\footnotesize
        \State Receive load demand $\bD[n]$
        \State Compute system state $\bX[n]$ (e.g., by power-flow simulations)
        \If{system is~\emph{critical}~\eqref{eq:riskthreshold}}\\
        \hspace*{0.4em}
        \textbf{\mbox{choose action $a[n] \in \mcA$ with probability $\epsilon_{n}$, run:}}
        \hspace*{2.6em}
        \Procedure{Physics-guided Explore}{}\\
        \hspace*{3.5em} Use $\bX[n]$ to refine action set $\mcA_{\sf line}$ to $\mcR^{\sf eff}_{\sf line}[n]$ (using Algorithm~\ref{alg:algoLineRemoval})\\
        \hspace*{3.5em} Select exploration action $a[n] \in \mcR^{\sf eff}_{\sf line}[n]\bigcup \mcA_{\sf gen}$ (using Algorithm~\ref{alg:algoExplore})
        \EndProcedure\\
        \hspace*{0.4em}
        \textbf{\mbox{otherwise (with probability $1-\epsilon_{n}$), run:}}
        \Procedure{$Q$-guided Exploit}{}\\
        %\hspace*{1.2em} \MakeUppercase{$Q$-guided Exploit} with probability $1 - \epsilon_{n}$\\
        \hspace*{2.5em} Use $\bX[n]$ to infer MDP state $\bS[n]$~\eqref{eq: mdp state}\\
        \hspace*{2.5em} Use MDP state $\bS[n]$ to predict $Q$-values using DQN\\     
        \hspace*{2.5em} Select exploitation action $a[n]\in \mcA$ based on $Q$-values (using Algorithm~\ref{alg:algoExploit})
        \EndProcedure
        \State Execute chosen action $a[n]$ in the power system simulator
        \State Update $\epsilon_{n}$ and the parameters of the DQN %$\btheta_{n}$ via stochastic gradient descent
        \EndIf
        %\State $\trainSteps \leftarrow \trainSteps + 1$
        \If{$n < T\;\&\;\mbox{No Blackout}$}
            \State $n \leftarrow n + 1$
            \State \mbox{Return to Step (1)} 
        %\Else
        %    \State $n \leftarrow 1$
        %    \State \mbox{Start new scenario}
        \EndIf
    \end{algorithmic}
\end{algorithm}

\begin{comment}
\subsection{Framework Overview}
A meta algorithm goes here
\begin{enumerate}
    \item First rethink Algorithm 2 and rewrite it (follow board discussions)
    \item Algorithm 1
    \item Algorithm 3
    \item Algorithm 4
\end{enumerate}
\end{comment}

\subsection{Learning an Effective Remedial Control Policy $\pi$}

Algorithm~\ref{alg:train} outlines the $Q$-learning procedure for learning a policy $\pi$. We employ a fully-connected neural network (NN), parameterized by $\btheta$, to predict $Q$-values~\eqref{eq: value function}. The NN is succinctly referred to as the deep $Q$-network (DQN$_{\btheta}$), and the resulting policy is denoted by $\pi_{\btheta}$. At time $n$, the agent receives load demand $\bD[n]$ and infers $\bX[n]$ by power-flow simulations. Next, an action $a[n] \in \mcA$ is chosen, contingent on whether the system is~\emph{critical} according to~\eqref{eq:riskthreshold}. If not critical, the agent ``reconnects" the line that maximizes the reward~\emph{estimate} $\tilde r[n]\in \mcR$ if legally viable; otherwise, it chooses the ``do-nothing" action. When the system is critical, however, the agent employs an $\epsilon_{n}$-greedy exploration scheme to choose an action $a[n]\in\mcA$, i.e., the agent chooses to explore with probability $\epsilon_{n}$ via Algorithm~\ref{alg:algoExplore} and exploit with probability $1-\epsilon_{n}$ via Algorithm~\ref{alg:algoExploit}. The selected action $a[n]\in\mcA$ is executed in the power system simulator, followed by an updating $\btheta_{n}$ to improve policy $\pi_{\btheta_{n}}$. Next, we describe the exploration and exploitation procedures.

\paragraph{\bf Physics-Guided Exploration (Algorithm~\ref{alg:algoLineRemoval} and \ref{alg:algoExplore})} Initially, the agent lacks information about the system dynamics and thus, relies on prior knowledge to choose exploratory actions~$a\in \mcA$. We employ sensitivity factors to guide exploration, based on the following key idea. Topological actions that~\emph{decrease} line flows below line limits $F_{\ell}^{\sf max}$ without causing overloads in~\emph{other healthy} lines $\ell \in \mcL[n]$ ensure a transition to a better system state in the short term. This approach ensures a transition to MDP states that may otherwise be challenging to reach by taking {random} control actions.

\begin{algorithm}[t]
    \caption{Construct Set $\mcR^{\sf eff}_{\sf line}[n]$ from Action Space $\mcA^{}_{\sf line}$}
    \label{alg:algoLineRemoval}
    \begin{algorithmic}[1]
        %\scriptsize
        %\footnotesize
        %\tiny
        \Procedure{Effective Set $\mcR^{\sf eff}_{\sf line}$}{$\mcA^{}_{\sf line}$} 
            \State Observe system state $\bX[n]$ and construct $\mcL[n]$
            \State Initialize $\mcR^{\sf eff}_{\sf line}[n] \leftarrow \emptyset$
            \State Construct $\mcA^{\sf rem}_{\sf line}[n] \leftarrow \{\ell \in \mcL[n]: \tau_{\rm D} = 0\;\mbox{\&}\;\tau_{\rm F} = 0\}$~\Comment{\textcolor{red}{legal removals}}
            \State Construct $\ptdf[n] \in \R^{L \times N}$ matrix from $\bX[n]$ 
            \State Construct $\lodf[n] \in \R^{L \times L}$ matrix from $\ptdf[n]$
            \State Find $\ell_{\sf max}=\dff \argmax_{\ell \in \mcL[n]} \;\; \rho_{\ell}[n]$
            \For {line $k$ in $\mcA^{\sf rem}_{\sf line}[n]\backslash \{\ell_{\sf max}\}$}~\Comment{\textcolor{red}{legal line removals that decrease flow}}
                \State Compute $F_{\ell_{\sf max}}[n+1] \leftarrow F_{\ell_{\sf max}}[n] + \lodf_{\ell_{\sf max}, k} \cdot F_{k}[n]$ 
                    \If{$|F_{\ell_{\sf max}}[n+1]| \leq F^{\sf max}_{\ell_{\sf max}}$}
                        \State $\mcR^{\sf eff}_{\sf line}[n] \leftarrow \mcR^{\sf eff}_{\sf line}[n] \bigcup \{k\}$
                    \EndIf		
            \EndFor
            \For {line $k$ in $\mcR^{\sf eff}_{\sf line}[n]$}~\Comment{\textcolor{red}{no additional overloads}}
                \For {line $\ell$ in $\mcL[n] \backslash \{\ell_{\sf max}\}$} 
                    \State Compute $F_{\ell}[n+1] \leftarrow F_{\ell}[n] + \lodf_{\ell, k} \cdot F_{k}[n]$ 
                    \If{$|F_{\ell}[n+1]| > F^{\sf max}_{\ell}$}
                        \State $\mcR^{\sf eff}_{\sf line}[n] \leftarrow \mcR^{\sf eff}_{\sf line}[n] \backslash \{k\}$
                        \State~\textbf{Break}
                    \EndIf
                \EndFor
            \EndFor
            \State Construct $\mcA^{\sf reco}_{\sf line}[n] \leftarrow \{\ell \in \neg \mcL[n]: \tau_{\rm D} = 0\;\mbox{\&}\;\tau_{\rm F} = 0\}$~\Comment{\textcolor{red}{legal reconnect}}
            \State $\mcR^{\sf eff}_{\sf line}[n] \leftarrow \mcR^{\sf eff}_{\sf line}[n] \bigcup \mcA^{\sf reco}_{\sf line}[n]$	
            \State \Return $\mcR^{\sf eff}_{\sf line}[n]$
    \EndProcedure
    \end{algorithmic}
\end{algorithm}

Identifying actions based on sensitivity factors and flow models~\eqref{eq: lodf} faces the challenge that removing a line $k\in\mcL[n]$ can simultaneously reduce flow in certain lines while increasing flow in others. To rectify this, we focus on identifying remedial actions that reduce flow in the currently~\emph{maximally} loaded line. At time $n$, we define the maximally loaded line index $\ell_{\sf max}\dff \argmax_{\ell \in \mcL[n]} \;\; \rho_{\ell}[n]$. 
% as
% \begin{equation}
% \begin{aligned}
%     \label{eq: max line index}
%         \ell_{\sf max} \dff \argmax_{\ell \in \mcL[n]} \;\; \rho_{\ell}[n]\ .
% \end{aligned}
% \end{equation} 
Leveraging the structure of the $\lodf[n]$ matrix, we design an~\emph{efficient} algorithm for identifying an~\emph{effective} set $\mcR^{\sf eff}_{\sf line}[n]$ consisting of actions $a[n] \in \mcA_{\sf line}$ that greedily decrease risk margins $\rho_{\ell_{\sf max}}[n]$. Specifically, Algorithm~\ref{alg:algoLineRemoval} has three main steps. First, the agent constructs a legal action set $\mcA^{\sf rem}_{\sf line}[n] \subset \mcA_{\sf line}$ from $\bX[n]$, comprising of permissible line removal candidates. Specifically, lines $\ell \in \mcL[n]$ with legality conditions $\tau_{\rm D} = 0$ and $\tau_{\rm F} = 0$ can only be removed rendering other control actions in $\mcA_{\sf line}$ irrelevant at time $n$. Second, a dynamic set $\mcR^{\sf eff}_{\sf line}[n]$ is constructed by initially identifying lines $k\in \mcA^{\sf rem}_{\sf line}[n]\backslash \{\ell_{\sf max}\}$ whose removal~\emph{decrease} flow in line $\ell_{\sf max}$ below its rated limit $F^{\sf max}_{\ell_{\sf max}}$. Finally, the agent eliminates lines from $\mcR^{\sf eff}_{\sf line}[n]$ the removal of which creates additional overloads in the network. Note that we include~\emph{all} currently disconnected lines $\ell \in \neg \mcL[n]$ as potential candidates for reconnection in the set $\mcR^{\sf eff}_{\sf line}[n]$, provided they adhere to legality conditions ($\tau_{\rm D} = 0$ and $\tau_{\rm F} = 0$). It is noteworthy that the set $\mcR^{\sf eff}_{\sf line}[n]$ is~\emph{time-varying}. Hence, depending on the current system state $\bX[n]$, $\mcR^{\sf eff}_{\sf line}[n]$ may either contain a few elements or be empty.

Next, the agent aims to select an exploratory action $a[n]\in \mcA$ with probability $\epsilon_{n}$, guided by our designed physics-guided exploration policy outlined in Algorithm~\ref{alg:algoExplore}. This policy involves constructing the set $\mcR^{\sf eff}_{\sf line}[n]$ from Algorithm~\ref{alg:algoLineRemoval}. Subsequently, as each action $a \in \mcR^{\sf eff}_{\sf line}[n] \cup \mcA_{\sf gen}$ is potentially effective, the agent chooses an action $a[n]$ that maximizes the reward estimate $\tilde r[n]$. The reward estimate is calculated using the flow models~\eqref{eq: ptdf comb} or~\eqref{eq: lodf} based on the considered action.

\begin{algorithm}[t] 
	\caption{Physics-Guided \emph{Exploration} with Probability $\epsilon_{n}$}
	\label{alg:algoExplore}
	\begin{algorithmic}[1]
            %\scriptsize
            %\footnotesize
		\Procedure{Physics-guided Explore}{$\mcA_{\sf line}, \mcA_{\sf gen}$}
            \State Construct $\mcR^{\sf eff}_{\sf line}[n]$ from $\mcA_{\sf line}$ using Algorithm~\ref{alg:algoLineRemoval}
		\State Initialize $\maxReward \gets -\infty$
		\State Initialize $\maxAction \gets$ \textbf{None}
		\For {each action $a$ in $\mcR^{\sf eff}_{\sf line}[n] \bigcup \mcA_{\sf gen}$}~\Comment{\textcolor{red}{get reward estimate}}
		\State Obtain reward estimate $\tilde r[n]$ for action $a$ via flow models~\eqref{eq: ptdf comb} or~\eqref{eq: lodf} 
		\If{$\tilde r[n] > \maxReward$}
		\State $\maxReward \gets \tilde r$
		\State $\maxAction \gets a$
		\EndIf
		\EndFor
		\State \Return $\maxAction$
		\EndProcedure
	\end{algorithmic}
\end{algorithm}

\begin{algorithm}[t]
	\caption{$Q$-Guided \emph{Exploitation} with Probability $1 - \epsilon_{n}$}
	\label{alg:algoExploit}
	\begin{algorithmic}[1]
            %\scriptsize
            %\footnotesize
		\Procedure{$Q$-guided Exploit}{$\mcA_{\sf line}, \mcA_{\sf gen}, \btheta_{n}$}
            \State Infer MDP state $\bS[n]$ from $\bX[n]$
            \State Construct $\mcA^{\sf legal}_{\sf line}[n] \leftarrow \{\ell \in [L]: \tau_{\rm D} = 0\;\mbox{\&}\;\tau_{\rm F} = 0\}$~\Comment{\textcolor{red}{legal line-switch}}
		\State $\mcA^{\sf legal} \leftarrow \mcA^{\sf legal}_{\sf line}[n] \bigcup \mcA_{\sf gen}$
  		\State Initialize $\bQ[n] \leftarrow$ DQN$_{\btheta_{n}}(\bS[n])$ via~\eqref{eq: Q predict}	
		\State $\bQ_{\mcA^{\sf legal}}[n] \leftarrow \text{Filter}(\bQ[n], \mcA^{\sf legal})$\Comment{\textcolor{red}{filter \emph{legal} $Q$-values}}
		\State $\topFiveActions \leftarrow \text{TopFive}(\bQ_{\mcA^{\sf legal}}[n])$\Comment{\textcolor{red}{find top-$5$ legal $Q$-values}}
		\State Initialize $\maxReward \gets -\infty$
		\State Initialize $\maxAction \gets$ \textbf{None}
		\For {each action $a$ in $\topFiveActions$}~\Comment{\textcolor{red}{get reward estimate}}
		      \State Obtain reward estimate $\tilde r[n]$ for action $a$ via flow models~\eqref{eq: ptdf comb} or~\eqref{eq: lodf} 
			\If{$\tilde r[n] > \maxReward$}
			\State $\maxReward \gets \tilde r[n]$
			\State $\maxAction \gets a$
			\EndIf
		\EndFor
		\State \Return $\maxAction$
		\EndProcedure
	\end{algorithmic}
\end{algorithm}

\paragraph{\bf $Q$-Guided Exploitation Policy (Algorithm~\ref{alg:algoExploit})} 
The agent refines its action choices over time by leveraging the feature representation $\btheta_{n}$, learned through the minimization of the temporal difference error via stochastic gradient descent. Specifically, the agent employs the current DQN$_{\btheta_{n}}$ to select an action~$a\in \mcA$ with probability $1 - \epsilon_{n}$. The process begins with the agent inferring the MDP state $\bS[n]$ in~\eqref{eq: mdp state} from $\bX[n]$. Next, the agent predicts a $\bQ[n]\in \R^{|\mcA|}$ vector using the network model DQN$_{\btheta_{n}}(\bS[n])$ through a forward pass, where each element represents $Q$-value predictions associated with each remedial control actions $a[n]\in \mcA$. Rather than choosing the action with the highest $Q$-value, the agent first identifies legal action subset $\mcA^{\sf legal} \dff \mcA^{\sf legal}_{\sf line}[n] \bigcup \mcA_{\sf gen}$ from $\bX[n]$. Next, the agent identifies actions $a[n]\in \mcA^{\sf legal}$ associated with the top-$5$ $Q$-values within this legal action subset $\mcA^{\sf legal}$ and chooses one optimizing for the reward estimate $\tilde r[n]$ (computed via~\eqref{eq: ptdf comb} and~\eqref{eq: lodf}). This policy 
%mitigates the challenge of choosing illegal actions, 
accelerates learning without the need to design a sophisticated reward function $\mcR$ that penalizes illegal actions.

\paragraph{\bf{DQN Architecture}} Our DQN architecture consists of a fully-connected NN, parameterized by $\btheta$, that takes as input a critical MDP state $\bS[n]$ and~\emph{predicts} two key quantities of interest: (i) a state-value function $V_{\btheta}(\bS[n])$ and, (ii) an advantage function $A_{\btheta}(\bS[n], a[n])$ that quantifies the advantage of taking action $a[n]$ in MDP state $\bS[n]$. $Q$-values~\eqref{eq: value function} for each action $a[n]\in \mcA$ are then predicted via
\begin{equation}
\begin{aligned}
    \label{eq: Q predict}
        Q_{\btheta}(\bS[n], a[n]) \dff A_{\btheta}(\bS[n], a[n]) + V_{\btheta}(\bS[n])\ .
\end{aligned}
\end{equation} 
This~\emph{dueling} NN architecture~\cite{wang2016dueling} takes $\bS[n]$ as it's input, denoted by DQN$_{\btheta}(\bS[n])$, and outputs an advantage vector $\bA_{\btheta}(\bS[n], \cdot)\in \R^{|\mcA|}$ along with a state-value function $V_{\btheta}(\bS[n])$ implicitly predicting $\bQ_{\btheta}(\bS[n],\cdot) \in \R^{|\mcA|}$ via~\eqref{eq: Q predict}.

\paragraph{Experience Replay}
To enhance the learning efficiency of $\pi_{\btheta}$ via DQN$_{\btheta}$, we employ a~\emph{prioritized} experience buffer~\cite{Schaul:2016} that stores past experience according to a~\emph{priority} encountered during the agent's interaction with the power system. The past experience is captured by the tuples $(\bS[n], a[n], r[n], \bS[n+1], \isend(\bS[n+1]))$, where $\isend(\bS)$ is a boolean value such that when it is true, it indicates that MDP state $\bS$ is the~{terminal} state. The prioritization of past experience tuples, based on the magnitude of temporal-difference errors, when sampling mini-batches for DQN$_{\btheta}$ training contributes to more effective learning of behavior strategies $\pi_{\btheta}$ influenced by~$\btheta$~\cite{Schaul:2016}.

\paragraph{Training the DQN$_{\btheta}$}
To ensure training stability, we employ the standard approach of splitting the task of predicting and evaluating $Q$-values via two separate NNs~\cite{Mnih:2015}: a behavior DQN$_{\btheta}$ and a target DQN$_{\btheta^{-}}$ parameterized by a distinct set of parameters $\btheta$ and $\btheta^{-}$, respectively. Under critical states $\bS[n]$, the agent samples $B$ mini-batches of experience tuples, each of type $(\bS[i], a[i], r[i], \bS[i+1], \isend(\bS[i+1]))$, from the replay buffer on which the target DQN$_{\btheta^{-}}$ is~\emph{unrolled} to predict $Q$-values~\eqref{eq: Q predict}. Subsequently, $Q$-values are employed to compute a~\emph{look-ahead} bootstrapped target $t[i]$ for each mini-batch via
\begin{equation}
\begin{aligned}
    \label{eq: target update}
        t[i] = r[i] +  \gamma \cdot (1 - \isend(\bS[i+1])) \cdot \max_{a}Q_{\btheta^{-}}(\bS[i+1], a)\ ,
\end{aligned}
\end{equation} which is further used to update the behavior DQN$_{\btheta}$ network parameters via a stochastic gradient descent update
\begin{align}
    \label{eq: gradient update}
    \btheta_{n+1} & = \btheta_{n} - \alpha_{n} \cdot \nabla_{\btheta_{}} 
    \left[\;t[i] - Q_{\btheta}(\bS[i], a[i])\;\right]^{2}\ , 
\end{align} where $\alpha_{n}$ denotes the learning rate and $n$ denotes the current training iteration. We use the Adam optimizer~\cite{adam:2015} to perform the gradient update~\eqref{eq: gradient update}, an inverse-time decay schedule for $\alpha_{n}$, and update the target DQN$_{\btheta^{-}}$ parameters $\btheta^{-} = \btheta$ by following a~\emph{soft} parameter update scheme every iteration.

\section{Case Studies and Discussion}
\label{sec: Case Studies and Discussion}

%\subsection{Simulation Platform and Dataset}
We assess the effectiveness of our framework by using Grid2Op~\cite{donnot:2020}, an open-source gym-like platform simulating power transmission networks with real-world operational constraints.
Simulations run on a Grid2Op test network, consisting of a diverse dataset with various scenarios. Each scenario encompasses generation $\bG[n]$ and load demand $\bD[n]$ set-points for all time steps $n\in[T]$ across every month throughout the year. Each scenario represents approximately 28 days with a 5-minute time resolution, based on which we have horizon $T=8062$. We train and test several agents on the Grid2Op 36-bus and {the IEEE 118-bus power network,} and use TensorFlow~\cite{tensorflow2015} for training the behavior and target DQNs. 

Grid2Op offers diverse scenarios throughout the year with distinct monthly load profiles. 
December consistently shows high aggregate demand, pushing transmission lines closer to their maximum flow limits while May experiences lower demand.  {We perform analysis on the Grid2Op 36-bus and IEEE 118-bus systems. For both systems, we have performed a~\emph{random} split of Grid2Op scenarios. For the test sets, we selected 32 scenarios for the Grid2Op 36-bus system and 34 scenarios for the IEEE 118-bus system, while assigning 450 scenarios to the training sets and a subset for validation to determine the hyperparameters, as discussed in Section~\ref{sec: Parameters and Hyperparameters}.} 
%we have performed a~\emph{random} split of Grid2Op scenarios. We have chosen $32$ scenarios for the test set, $450$ for the training set, and a subset for validation, which is used to determine hyperparameters, as discussed in Section~\ref{sec: Parameters and Hyperparameters}. \textcolor{blue}{Similarly, for the IEEE 118-bus system,  we have chosen $34$ scenarios for the test set, assigned $450$ to the training set, and set aside a subset for validation.}
To ensure proper representation of various demand profiles, the test set includes at least two scenarios from each month.
\begin{table}[t]
	\centering
  \scalebox{.8}{
		\begin{tabular}{|c | c | c | c|} 
			\hline
			System-State Feature $\bX[n]$ & Size & Type & Notation \\ [0.4ex] 
			\hline\hline
			\texttt{prod\_p} & $G$ & float & $\bG[n]$ \\ 
			\hline
			\texttt{load\_p} & $D$ & float & $\bD[n]$ \\ 
			\hline
			\texttt{p\_or, p\_ex} & $L$ & float & $F_{\ell}[n]$ \\ 
			\hline
			\texttt{a\_or, a\_ex} & $L$ & float & $A_{\ell}[n]$ \\ 
			\hline
			\texttt{rho} & $L$ & float & $\rho_{\ell}[n]$ \\ 
			\hline
			\texttt{line\_status} & $L$ & bool & $\mcL[n]$ \\ 
			\hline
			\texttt{timestep\_overflow} & $L$ & int & overload time \\ 
			\hline
			\texttt{time\_before\_cooldown\_line} & $L$ & int & line downtime \\ 
			\hline
			\texttt{time\_before\_cooldown\_sub} & $N$ & int & bus downtime \\ 
			\hline
		\end{tabular}
	}
	\caption{Heterogeneous input system state features $\bX[n]$.}
	\label{tab:inputFeatures}
\end{table}

%\vspace{-0.1in}

\subsection{Parameters and Hyper-parameters}
\label{sec: Parameters and Hyperparameters}

%\paragraph{\bf System's Parameters}
\noindent {\bf System's Parameters:} The Grid2Op 36-bus system consists of $N=36$ buses, $L=59$ transmission lines (including transformers), $G=10$~\emph{dispatchable} generators, and  $D=37$ loads. We employ $F=8$ line and $H=3$ bus features (Table~\ref{tab:inputFeatures}), totaling $O =567$~\emph{heterogeneous} input system state $\bX[n]$ features. Each MDP state $\bS[n]$ considers the past $\kappa = 6$ system states for decision-making. Without loss of generality, we set $\eta=0.95$ and $\nu=0$ specified  in~\eqref{eq:riskthreshold} as the thresholds for determining whether the system is in critical or normal states, respectively. 
{The IEEE 118-bus system consists of $N=118$ buses, $L=186$ transmission lines, 
%(including transformers), 
$G=32$~\emph{dispatchable} generators, and $D=99$ loads. While in principle we can choose all the 11 features, to improve the computational complexity we choose a subset of line-related features, specifically, $F=5$ line features ($\texttt{p\_or, a\_or, rho, line\_status}$ and $\texttt{timestep\_overflow}$). This results in a total of $O = 930$~\emph{heterogeneous} input system state features and consider the past $\kappa = 5$ system states for decision-making. Without loss of generality, we set $\eta=1.0$ and $\nu=0$.}

After performing a line-switch action $a[n]\in\mcA_{\sf line}$ on any line $\ell$, we impose a mandatory downtime of $\tau_{\rm D}=3$ time steps (15-minute interval) for each line $\ell \in [L]$. In the event of natural failure caused due to an overload cascade, we extend the downtime to $\tau_{\rm F}=12$ (60-minute interval).

\noindent  {\bf DQN Architecture and Training:} Our DQN architecture features a feed-forward NN with two hidden layers, each having $O$ units and adopting~\texttt{tanh} nonlinearities. The input layer, with a shape of $|\bS[n]| = O \cdot \kappa$, feeds into the first hidden layer of $O$ units, followed by another hidden layer of $O$ units. The network then splits into two streams: an advantage-stream $\bA_{\btheta}(\bS[n], \cdot)\in \R^{|\mcA|}$ with a layer of $|\mcA|$ action-size units and~\texttt{tanh} non-linearity, and a value-stream $V_{\btheta}(\bS[n]) \in \R$ predicting the value function for the current MDP state $\bS[n]$. $\bQ_{\btheta}(\bS[n], \cdot)$ are obtained by adding the value and advantage streams~\eqref{eq: Q predict}. We penalize the reward function $r[n]$ in~\eqref{eq: reward} in the event of failures attributed to overloading cascades and premature scenario termination ($n < T$). 
%Additionally, we normalize the reward constraining its values to the interval $[-1, 1]$. 
For the Grid2Op 36-bus system, we use a learning rate $\alpha_{n}=5\cdot 10^{-4}$ decayed every $2^{10}$ training iterations, a mini-batch size of $B=64$, an initial $\epsilon = 0.99$ exponentially decayed to $\epsilon = 0.05$ over $26\cdot 10^{3}$ agent-MDP training interaction steps and choose $\gamma = 0.99$. {Likewise, for the IEEE 118-bus system we use similar parameters with a mini-batch size of $B=32$.}
%{Likewise, for the IEEE 118-bus system we set $\alpha_{n}=9\cdot 10^{-4}$ with a mini-batch size of $B=32$ and $21\cdot 10^{3}$ agent MDP training interaction steps.}

\vspace{0.05 in}
\noindent {\bf Line-Switch Action Space Design $\mcA_{\sf line}$:} For the considered system, following the MDP modeling discussed in Section~\ref{sec: Action Space}, for the Grid2Op 36-bus system we have $|\mcA_{\sf line}|=119\;(2L + 1)$ and {for the IEEE 118-bus system we have  $|\mcA_{\sf line}|=373\;(2L + 1)$.} 
%Without loss of generality, 
We set a uniform line-switch cost $c^{\sf line}_{\ell}=1$ for all lines $\ell \in [L]$.

\begin{table}[h]
    \centering
    \scalebox{.8}{
        \begin{tabular}{|c | c | c | c|} 
            \hline
            \thead{Generator}		
            &\thead{Ramp-Rate \\ $\Delta \bG^{\sf max}_{j}$ (in MWs)}  
            &\thead{Maximum \\ Generation $\bG^{\sf max}_{j}$} 	
            &\thead{Cost $c^{\sf gen}_{j}$ \\ (in $\$$/MW)}   \\ [0.4ex] 
            \hline\hline
            $G_{1}$ 		& $10.4$ & $250$ & $36$ \\ 
            \hline
            $G_{2}$  		& $9.9$  & $350$ & $40$ \\ 
            \hline
            $G_{3}$  		& $8.5$	 & $300$ & $48$ \\ 
            \hline
            $G_{4}$  		& $4.3$  & $150$ & $46$ \\ 
            \hline
            $G_{5}$ 		& $2.8$  &$100$    & $44$  \\ 
            \hline
        \end{tabular}
    }
    \caption{Generator features for the Grid2Op 36-bus network.}
    \label{tab:genChars}
\end{table}

\noindent {\bf Generator Action Space Design $\mcA_{\sf gen}$:} For the Grid2Op 36-bus system, we select a subset consisting of $k=5$ dispatchable generators (out of $G$) with the largest ramp-rates $\Delta G^{\sf max}_{j}$ to design $\mcA_{\sf gen}$. The characteristics of the chosen generators are shown in Table~\ref{tab:genChars}. We choose a discretization constant $\delta = 2\; \leq \min_j \Delta G_{j}^{\sf max} $ to ensure all the selected generators actively participate in the decisions ensuring that the ramp-rate constraints are satisfied for all generators. Since each feasible generator adjustment combination~\eqref{eq: generator action} is designed as an action $a[n]\in \mcA_{\sf gen}$, the total number of actions for a given choice of $\delta$ is calculated via dynamic programming, rendering $|\mcA_{\sf gen}| = 50$. {For the IEEE 118-bus system, we select a subset consisting of $k=6$ dispatchable generators with the largest ramp-rates $\Delta G^{\sf max}_{j}$ to design $\mcA_{\sf gen}$ and choose a discretization constant $\delta = 5\; \leq \min_j \Delta G_{j}^{\sf max} $, rendering $|\mcA_{\sf gen}| = 140$.}

%\vspace{-0.15in}
\subsection{Evaluation Criteria}
\label{sec: Evaluation Criteria}

\noindent {\bf Performance Metrics:} A key performance metric is the agent's survival time ST$(T)$, averaged across all test set scenarios for $T=8062$. We explore factors influencing ST through analyzing action diversity and track~\emph{unique} control actions per scenario. Furthermore, we quantify the fraction of times each of the following three possible actions are taken: ``do-nothing," ``line-switch $\mcA_{\sf line}$," and ``generator adjustment $\mcA_{\sf gen}$". {Since the agent takes remedial actions only under critical states associated with critical time instances $n$, we report action decision fractions that exclusively stem from these critical states, corresponding to instances when $\rho_{\ell_{\sf max}}[n] \geq \eta$. We also note that monthly load demand variations $\bD[n]$ influence how frequently different MDP states $\bS[n]$ are visited. This results in varying control actions per scenario. To form an overall insight, we report the \emph{average} percentage of actions chosen across all test scenarios. Note that the reward function $r[n]$ in~\eqref{eq: reward} is characterized by its dependence on parameters $(\mu_{\sf gen}, \mu_{\sf line})$, designed to accommodate control action costs $\beta_{\sf line}$ and $\beta_{\sf gen}$ associated with line-switch and generator adjustments, respectively. Consequently, the choice of $(\mu_{\sf gen}, \mu_{\sf line})$ directly influences performance metrics. To illustrate the benefits of physics-guided exploration, we showcase how performance metrics vary as a function of $(\mu_{\sf gen}, \mu_{\sf line})$, providing a comprehensive understanding of the agent's adaptive decision-making for different cost scenarios.

\vspace{.05 in}
\noindent {\bf Baseline Agents:} For the chosen performance metrics, we consider~{five} alternative baselines: (i)~$\noop$ agent consistently opts for the ``do-nothing" action across all scenarios, independent of the system-state $\bX[n]$; (ii) $\reco$ agent decides to ``re-connect" a disconnected line that greedily~\emph{maximizes} the reward estimate $\tilde r[n]$~\eqref{eq: reward} at the current time step $n$. In cases where reconnection is infeasible due to line downtime constraints or when no lines are available for reconnection, the $\reco$ agent defaults to the ``do-nothing" action for that step;~{(iii)~\texttt{milp\_agent}\cite{MILPAGENT:2022} agent strategically minimizes over-thermal line margins using line switching actions $\mcA_{\rm line}$ by formulating the problem as a mixed-integer linear program (MILP); (iv)~\texttt{OptimCVXPY}\cite{OptimCVXPY:2022} agent redispatches the generators to proactively minimize thermal limit margins by formulating a continuous optimization problem. For comparisons, agents~\texttt{milp\_agent} and \texttt{OptimCVXPY} are activated only under critical states, i.e., when $\rho_{\ell_{\sf max}}[n] \geq \eta$;} and (v)~RL + Random Explore baseline agent: we employ a DQN$_{\btheta}$ network with a~\emph{tailored} random $\epsilon_{n}$-greedy exploration policy during agent~{training}. Specifically, similar to Algorithm~\ref{alg:algoExplore}, the agent first constructs a legal action set $\mcA^{\sf legal}_{\sf line}[n] \dff \{\ell \in [L]: \tau_{\rm D} = 0 , \tau_{\rm F} = 0\}$ from $\bX[n]$ at critical times. In contrast to Algorithm~\ref{alg:algoExplore}, however, this agent chooses a~\emph{random} legal action in the set $a[n] \in \mcA^{\sf legal}_{\sf line}[n] \bigcup \mcA_{\sf gen}$ (instead of using $\mcR^{\sf eff}_{\sf line}[n]$). In the Grid2Op 36-bus system, using this random exploration policy, we train the DQN$_{\btheta}$ for $20$ hours of repeated interactions with the Grid2Op simulator for each $\mu_{\sf line} \in \{0, 0.5, 1, 1.5\}$. We report results associated with the~\emph{best} model $\btheta$ and refer to the best policy obtained following this random $\epsilon_{n}$-greedy exploration by $\pi^{\sf rand}_{\btheta}(\mu_{\sf line})$. {Similarly, in the IEEE 118-bus system, we train the DQN$_{\btheta}$ model for 15 hours of repeated interactions for $\mu_{\sf line}=0$. We present the results for the Grid2Op 36-bus system in the next two subsections and present their counterpart results to verify scalability in Subsection~\ref{sec:118}.} 
%\ATAD{mention the counterparts of the previous few lines for the 118 system.}
}

\subsection{36-bus System: Effect of Agent Type on Survival Time}
\label{sec: Effect of Agent Type on Survival Time}

To precisely capture the distinct nature of topological line-switch actions $\mcA_{\sf line}$ and generator adjustment actions $\mcA_{\sf gen}$ on improving ST, we incrementally construct our action space. %Specifically, we first restrict our action space to $\mcA = \mcA_{\sf line}$ and subsequently, consider $\mcA = \mcA_{\sf line} \bigcup \mcA_{\sf gen}$.

\paragraph{\bf Effect of Line-Switch $\mcA_{\sf line}$ on ST} Since this subsection exclusively explores the effect of line-switch control actions on ST, we choose $\mcA = \mcA_{\sf line}$ and $\mu_{\sf gen}=0$. Naturally, $\mcA_{\sf gen} = \emptyset$. Table~\ref{tab:line} presents and compares the results, illustrating increased agent sophistication as we move down the table. Starting from baselines, we observe that the $\noop$ agent achieves an average ST of 4,733 time steps, which is $58\%$ of the time-horizon and the $\reco$ agent performs marginally better than the $\noop$ agent by allocating $0.10\%$ of its actions towards reconnection at critical times. Intuitively, having a greater number of operational lines in the network reduces the likelihood of overloads, contributing to an improved average ST of 4,743.

%{comment about milp agent}

For policy $\pi^{\sf rand}_{\btheta}(\mu_{\sf line}=0)$ we observe an average ST of 5,929 steps, a $25.2\%$ increase over the baselines. Notably, $67.35\%$ of the remedial control actions chosen are~\emph{line removals}, demonstrating the effectiveness of strategically removing transmission lines in improving ST. Additionally, for assessing the performance when $\mu_{\sf line} \neq 0$, consider the reward function $r[n]$ in~\eqref{eq: reward}. A higher $\mu_{\sf line}$ imposes penalties on line-switch actions, promoting fewer line-switching. These observations are summarized in Table~\ref{tab:line}.

\begin{table}[t]
  \centering
  \renewcommand{\arraystretch}{1.1} % Adjust the value as needed
  \scalebox{0.8}{
  \begin{tabular}{|c|c|c|c|c|c|}
    \hline
    {\thead{Agent Type}}
    & \thead{$\mu_{\sf line}$}
    & \thead{Avg. ST}
    & \thead{$\%$\\ Do-nothing}
    & \thead{$\%$\\ Reconnect}
    & \thead{$\%$\\ Removals} \\
    \hline
    \hline
    $\noop$ & $-$    &$4733.96$ &$100$      &$-$      &$-$      \\
    \hline
    $\reco$ & $0$    &$4743.87$  &$99.90$   &$0.10$    &$-$     \\
    \hline
    {\texttt{milp\_agent}} &${0}$  &${4062.62}$ &${12.05}$ &${1.70}$  &${86.24}$  \\
    \hline
    \hline
    \multirow{4}{*}{\rotatebox[origin=c]{90}{\makecell{RL\;+ \\ Random\\Explore }}} 
    & $0$       &$5929.03$ &$26.78$ &$5.85$ &$67.35$  \\
    \cline{2-6}
    & $0.5$     &$5624.31$ &$92.27$ &$1.10$ &$6.62$   \\
    \cline{2-6}
    & $1.0$     &$5327.06$ &$81.51$ &$0.28$ &$18.20$ \\
    \cline{2-6}
    & $1.5$     &$4916.34$ &$92.69$  &$0.01$  &$7.28$ \\
    \hline
	 \hline
    \multirow{4}{*}{\rotatebox[origin=c]{90}{\makecell{RL\;+ \\ Physics\\Guided\\Explore}}} 
    & $0$    &$\mathbf{6657.09}$ &$1.74$ &$7.66$ &$90.59$     \\
    \cline{2-6}
    & $0.5$  &$\mathbf{6861.40}$ &$22.12$ &$5.50$ &$72.37$    \\
    \cline{2-6}
    & $1.0$  &$\mathbf{6603.56}$ &$13.93$ &$7.00$ &$79.06$    \\
    \cline{2-6}
    & $1.5$  &$\mathbf{6761.34}$ &$46.53$ &$6.12$ &$47.34$   \\
    \hline
  \end{tabular}
  }
  \caption{Average survival time as a function of $\mu_{\sf line}$ for the line-switch action space $\mcA = \mcA_{\sf line}$ with $\mu_{\sf gen}=0$ and $\eta = 0.95$.}
  \label{tab:line}
\end{table}

\begin{figure}[t]
\centering
\includegraphics[width=0.8\linewidth]{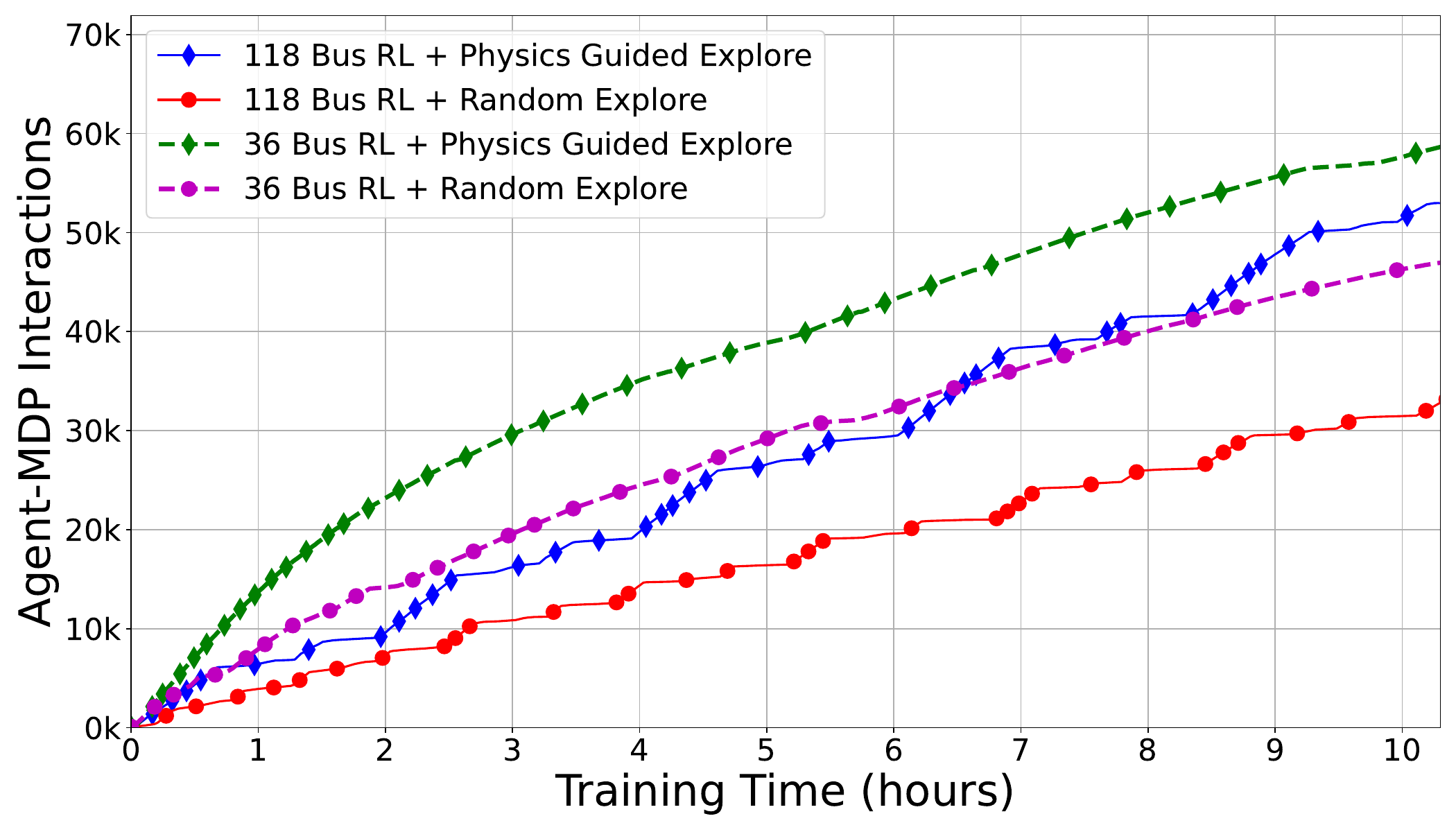}
\caption{{Agent$-$MDP interactions for both the Grid2Op 36-bus and the IEEE-118 bus systems with $\mcA = \mcA_{\sf line}$ and $\mu_{\sf line}=\mu_{\sf gen}=0$.}}
\label{fig:36bus_interactions}  
\end{figure}

Finally, we show the effectiveness of our physics-guided exploration with $\mcA_{\sf gen} = \emptyset$. We train the DQN$_{\btheta}$ following Algorithm~\ref{alg:train} and refer to the best policy obtained by $\pi^{\sf physics}_{\btheta}(\mu_{\sf line})$. To ensure fair comparisons with the random-exploration policy $\pi^{\sf rand}_{\btheta}(\mu_{\sf line})$, we train both DQN networks for the same duration of $20$ hours using the same hyperparameters reported in Section~\ref{sec: Parameters and Hyperparameters}. Starting from $\mu_{\sf line}=0$, we observe that the policy $\pi^{\sf physics}_{\btheta}$ achieves an average ST of 6,657 steps, representing a $40.6\%$ increase compared to baselines and a $12.2\%$ increase compared to $\pi^{\sf rand}_{\btheta}$. Notably, the physics-guided agent takes $25.05\%$ more line-switch actions compared to its random counterpart and can find more~\emph{effective} line-removal actions owing to the explicit design of $\mcR^{\sf eff}_{\sf line}[n]$ during agent training (Algorithm~\ref{alg:algoExplore}).~{To illustrate this effectiveness, Fig.~\ref{fig:36bus_interactions} shows the number of agent-MDP interactions as a function of training time. Despite the~\textit{more} computationally expensive update (due to the repeated $\lodf$ matrix computations) and~\textit{fewer} training epochs of $\pi^{\sf physics}_{\btheta}$, $\pi^{\sf physics}_{\btheta}$ achieves a~\textit{greater} number of agent-MDP interactions, indicating a more thorough state space exploration for a given computational budget. For example, by the end of the 14$^{\rm th}$ hour of training, $\pi^{\sf physics}_{\btheta}$ resulted in 68,492 agent-MDP interactions through 504 scenarios ($\approx$ 1.1 epochs), whereas $\pi^{\sf rand}_{\btheta}$ lead to 56,566 interactions through 1850 scenarios ($\approx$ 4.1 epochs).}

~{The ability of $\pi^{\sf physics}_{\btheta}$ to identify more effective actions, in comparison to $\pi^{\sf rand}_{\btheta}$, is further substantiated by incrementally increasing $\mu_{\sf line}$ and observing the performance changes. As $\mu_{\sf line}$ increases, the reward $r[n]$ becomes~\emph{less} informative about potentially effective actions due to the increasing penalties on line-switch actions, thus amplifying the importance of physics-guided exploration design. This is observed in Table~\ref{tab:line}, where unlike the policy $\pi^{\sf rand}_{\btheta}(\mu_{\sf line})$, the ST associated with $\pi^{\sf physics}_{\btheta}$ does not degrade as $\mu_{\sf line}$ increases.}
%Furthermore, unlike the policy $\pi^{\sf rand}_{\btheta}(\mu_{\sf line})$, the ST does not degrade with an increase in $\mu_{\sf line}$, underscoring the effectiveness of the physics-guided design. \AD{This is because} 
It is noteworthy that despite the inherent linear approximations of sensitivity factors, confining the RL exploration to actions derived from the set $\mathcal{R}^{\sf eff}_{\sf line}[n]$ enhances MDP state space exploration leading to $\pi^{\sf physics}_{\btheta}(\mu_{\sf line})$ consistently outperforming $\pi^{\sf rand}_{\btheta}(\mu_{\sf line})$.

\begin{table}[t]
  \centering
  \renewcommand{\arraystretch}{1.2} % Adjust the value as needed
  \scalebox{0.77}{
  \begin{tabular}{|c|c|c|c|c|c|c|}
      \hline
    {\thead{Agent Type}}
    & \thead{$\mu_{\sf line}$ } 
    & \thead{Avg. ST} 
    & \thead{$\%$\\ Do-nothing}
    & \thead{$\%$\\ Reconnect}
    & \thead{$\%$\\ Removals}
    & \thead{$\%$\\ Re-dispatch}  \\
    \hline
    \hline
    {\texttt{OptimCVXPY}} &${0}$  &${5471.84}$ &$-$ &$-$  &$-$  &${100}$      \\
    \hline
    \hline
    \multirow{4}{*}{\rotatebox[origin=c]{90}{\makecell{RL\;+ \\ Random\\Explore}}} 
    & $0$   &$6766.12$  &$0.65$ &$0.06$ &$47.05$ &$52.22$ \\
    \cline{2-7}
    & $0.5$ &$6630.59$  &$17.76$ &$1.32$ &$6.53$ &$74.37$  \\
    \cline{2-7}
    & $1.0$ &$6453.96$  &$5.08$ &$0.20$ &$16.94$ &$77.75$  \\
    \cline{2-7}
    & $1.5$ &$7218.25$  &$0.0$ &$0.02$ &$0.58$ &$99.39$  \\
    \hline
    \hline
    \multirow{4}{*}{\rotatebox[origin=c]{90}{\makecell{RL\;+ \\ Physics\\Guided\\Explore}}} 
    & $0$    &$\mathbf{7176.44}$    &$0.0$  &$2.52$ &$66.93$ &$30.55$  \\
    \cline{2-7}
    & $0.5$  &$\mathbf{7033.25}$    &$1.51$ &$0.94$ &$39.54$ &$58.01$  \\
    \cline{2-7}
    & $1.0$  &$\mathbf{6776.65}$    &$0.0$  &$0.90$ &$16.24$ &$82.85$   \\
    \cline{2-7}
    & $1.5$  &$\mathbf{7586.59}$    &$0.0$  &$0.31$ &$4.05$  &$95.64$  \\
    \hline
  \end{tabular}
  }
  \caption{Average survival time as a function of $\mu_{\sf line}$ for the hybrid action space $\mcA = \mcA_{\sf line} \bigcup \mcA_{\sf gen}$ with $\mu_{\sf gen}=0$ and $\eta = 0.95$.}
  \label{tab:lineAndGen}
\end{table}

% \begin{figure}[t]
% \centering
% \includegraphics[width=0.75\linewidth]{Cascading Failure Mitigation/Figures/36bus_EnvInteractions.pdf}
% \caption{{Agent$-$MDP interactions for the Grid2Op 36-bus system with $\mcA = \mcA_{\sf line}$ and $\mu_{\sf line}=\mu_{\sf gen}=0$ for $\eta = 0.95$.}}
% \label{fig:36bus_interactions}  
% \end{figure}

\paragraph{\bf Effect of Generator Adjustments $\mcA_{\sf gen}$ on ST} Next, we illustrate that strategic line-switching when combined with multiple generator adjustments further improves ST, showcasing the merits of our generator action space design $\mcA_{\sf gen}$. To this end, we augment our action space to $\mcA = \mcA_{\sf line} \bigcup \mcA_{\sf gen}$. Along with~$\noop$ and $\reco$ agents, we consider DQN$_{\btheta}$ agents, both with random and physics-guided exploration, in Table~\ref{tab:line} as baselines. To make the comparisons with Table~\ref{tab:line} fair, we train all the agents in this section with the same hyper-parameters reported in Section~\ref{sec: Parameters and Hyperparameters} and the same training time as in Section~\ref{sec: Evaluation Criteria}. We remark that although $\mu_{\sf gen}=0$ was chosen for fair comparisons with Table~\ref{tab:line}, any other desirable value for $\mu_{\sf gen}$ can be chosen too.  

The results are illustrated in Table~\ref{tab:lineAndGen}. Starting from $\mu_{\sf line}=0$, the agent with policy $\pi^{\sf rand}_{\btheta}(\mu_{\sf line}=0)$ survives 6,766 time steps on average, a $14.11\%$ increase compared to the agent with $\mcA_{\sf line}$ action space (Table~\ref{tab:line}). Importantly, $52.22\%$ of the actions chosen by the agent in critical states are generator adjustments, reducing the line-switch actions from $73.2\%$ to $47.11\%$. This illustrates that the combination of line-switch and multiple generator adjustments improves ST. When $\mu_{\sf line}\neq0$, we expect that a larger $\mu_{\sf line}$ encourages the agent to optimize for ST by taking more generator adjustment $\mcA_{\sf gen}$ actions. The observations are consistent with Table~\ref{tab:lineAndGen} where an increase in $\mu_{\sf line}$ leads to a consistent $\%$ increase in ``re-dispatch" actions. Note the drastically high average ST = 7,218 for policy $\pi^{\sf rand}_{\btheta}(\mu_{\sf line}=1.5)$, which is occurring despite a high penalty $\mu_{\sf line}$. This stems from the inherent challenge of precisely identifying factors that truly optimize ST. While our choice to prioritize risk margins in the reward function is evident, it may not explicitly capture all relevant factors, rendering the agent to optimize for unexpected behaviors.

Leveraging the physics-guided design in Algorithm~\ref{alg:train}, the agent associated with $\pi^{\sf physics}_{\btheta}(\mu_{\sf line}=0)$ survives 7,176 steps on average, a $6.06\%$ increase compared to $\pi^{\sf rand}_{\btheta}(\mu_{\sf line}=0)$. Importantly, the increase in performance can be attributed to a $22.34\%$ increase in line-switch (topological) actions compared to it's $\pi^{\sf rand}_{\btheta}(\mu_{\sf line})$ counterpart. When $\mu_{\sf line}\neq0$, similar to $\pi^{\sf rand}_{\btheta}(\mu_{\sf line})$, we observe an increase in the fraction of the ``re-dispatch" decisions. Furthermore, we observe that the policy $\pi^{\sf physics}_{\btheta}(\mu_{\sf line})$ always~{outperforms} all baselines for the same $\mu_{\sf line}$. Finally, similar to  $\pi^{\sf rand}_{\btheta}$, we observe that the agent $\pi^{\sf physics}_{\btheta}(\mu_{\sf line}=1.5)$ survives longest 7,586 time-steps on average ($94.10\%$ of the time-horizon), indicating that factors other than risk margins also contribute to improving ST.

%{comment about OptimCVXPY agent}

%This approach enables a comprehensive understanding of how the agent adapts its decision-making strategies under different cost scenarios.
%The last column reports the number of unique actions chosen by the agent (out of $|\mcA|=169$) averaged over the test set scenarios  
%It is noteworthy that the advantage of the setting $\kappa=3$ is viable at the expense of incurring a higher computational cost. This is due to more frequent weight updates rendering an inevitable accuracy-complexity trade-off.

\begin{table}[t]
  \centering
  \renewcommand{\arraystretch}{1.1} % Adjust the value as needed
  \scalebox{0.9}{
  \begin{tabular}{|c|c|c|c|c|c|c|}
    \hline
    {\thead{Agent Type}}
    & \thead{$\mu_{\sf line}$}
    & \thead{Avg. ST}
    & \thead{Avg. Action Diversity ($119$)}  \\
    \hline
    \hline
    $\noop$ & $-$    &$4733.96$ & $-$ \\
    \hline
    $\reco$ & $0$    &$4743.87$ &$1.093\;(1.821\%)$ \\
    \hline
    \hline
    {\texttt{milp\_agent}\cite{MILPAGENT:2022}} & $0$   &${4062.62}$ &${6.093\;(5.12\%)}$ \\
    \hline
    \hline
    \multirow{4}{*}{\rotatebox[origin=c]{90}{\makecell{RL\;+ \\ Random\\Explore}}} 
    & $0$       &$5929.03$ &$13.406\;(11.265\%)$ \\
    \cline{2-4}
    & $0.5$     &$5624.31$    &$4.218\;(3.544\%)$ \\
    \cline{2-4}
    & $1.0$     &$5327.06$ &$3.625\;(3.046\%)$ \\
    \cline{2-4}
    & $1.5$     &$4916.34$   &$3.406\;(2.862\%)$ \\
    \hline
    \hline
    \multirow{4}{*}{\rotatebox[origin=c]{90}{\makecell{RL\;+ \\ Physics\\Guided\\Explore}}} 
    & $0$    &$\mathbf{6657.09}$   &$\mathbf{17.062\;(14.337\%)}$ \\
    \cline{2-4}
    & $0.5$  &$\mathbf{6861.40}$   &$\mathbf{16.656\;(13.996\%)}$ \\
    \cline{2-4}
    & $1.0$  &$\mathbf{6603.56}$   &$\mathbf{17.156\;(14.416\%)}$ \\
    \cline{2-4}
    & $1.5$  &$\mathbf{6761.34}$   &$\mathbf{15.718\;(13.208\%)}$ \\
    \hline
  \end{tabular}
  }
  \caption{Action diversity as a function of $\mu_{\sf line}$ for the line-switch action space $\mcA = \mcA_{\sf line}$ with $\mu_{\sf gen}=0$ for $\eta = 0.95$.}
  \label{tab:lineDiversity}
\end{table}

\subsection{36-bus System: Effect of Agent Type on Action Diversity}
\label{sec: Effect of Agent Type on Action Diversity}

Beyond the percentage action split, the diversity of decisions across different MDP actions serves as an additional metric, providing insights into the impact of the exploration policy design on the agent's performance. Hence, we investigate the action diversity of agents after running all baselines and trained agents with different action spaces on our $32$ scenario test set.

First, we examine DQN$_{\btheta}$ agents with the line-switch action space $\mcA=\mcA_{\sf line}$. The results are presented in Table~\ref{tab:lineDiversity}. The agent policy $\pi^{\sf rand}_{\btheta}(\mu_{\sf line}=0)$ shows an average action diversity of $11.265\%$. This observation implies that the agent considers very few remedial line-switch control actions effective. Notably, a noteworthy trend emerges as $\mu_{\sf line}$ increases: the random exploration policy $\pi^{\sf rand}_{\btheta}(\mu_{\sf line})$ faces challenges in maintaining action diversity. Concurrently, a gradual decline in the average ST is also observed, underscoring the significance of action diversity in mitigating blackouts. This diminishing action diversity as $\mu_{\sf line}$ increases could be attributed to the agent's limited ability to thoroughly explore the MDP state space when following a random $\epsilon_{n}$-greedy exploration policy. In contrast, when compared to the physics-guided policy $\pi^{\sf physics}_{\btheta}(\mu_{\sf line}=0)$, we observe that the action diversity remains intact despite an expected decline in line-switches (Table~\ref{tab:line}). This is possible since the agent better explores the topological MDP state space more effectively owing to the agent's explicit interaction with the set $\mcR^{\sf eff}_{\sf line}[n]$ during training. 

Finally, we investigate DQN$_{\btheta}$ agents with the larger action space $\mcA=\mcA_{\sf line}\bigcup \mcA_{\sf gen}$. The outcomes are presented and compared in Table~\ref{tab:lineAndGenDiversity}, revealing two key observations. Firstly, the physics-guided policy $\pi^{\sf physics}_{\btheta}(\mu_{\sf line}=0)$ maintains an action diversity of $10.757\%$ nearly double than that of $\pi^{\sf rand}_{\btheta}(\mu_{\sf line}=0)$, coupled with a higher ST, emphasizing the advantages of physics-guided exploration. Secondly, as very few generator adjustment remedial control actions prove highly effective in improving ST, action diversity remains small and relatively stable as $\mu_{\sf line}$ increases, underscoring the full benefits of the meticulously designed action set $\mcA_{\sf gen}$.

\begin{table}[t]
  \centering
  \renewcommand{\arraystretch}{1.1} % Adjust the value as needed
  \scalebox{0.9}{
  \begin{tabular}{|c|c|c|c|c|c|c|c|}
      \hline
    {\thead{Agent Type}}
    & \thead{$\mu_{\sf line}$ } 
    & \thead{Avg. ST} 
    & \thead{Avg. Action Diversity ($169$)}  \\
    \hline
    \hline
    $\noop$ & $-$    &$4733.96$ &$-$ \\
    \hline
    $\reco$ & $0$    &$4743.87$ &$1.093\;(1.821\%)$ \\
    \hline
    \hline
    {\texttt{OptimCVXPY}\cite{OptimCVXPY:2022}} & $0$   &${5471.84}$ &${7.031\;(-)}$ \\
    \hline
    \hline
    \multirow{4}{*}{\rotatebox[origin=c]{90}{\makecell{RL\;+ \\ Random\\Explore}}} 
    & $0$ &$6766.12$       &$9.093\;(5.380\%)$ \\
    \cline{2-4}
    & $0.5$ &$6630.59$     &$7.468\;(4.418\%)$ \\
    \cline{2-4}
    & $1.0$ &$6453.96$     &$\mathbf{10.281\;(6.083\%)}$ \\
    \cline{2-4}
    & $1.5$ &$7218.25$     &$6.000\;(3.550\%)$ \\
    \hline
    \hline
    \multirow{4}{*}{\rotatebox[origin=c]{90}{\makecell{RL\;+ \\ Physics\\Guided\\Explore}}} 
    & $0$    &$\mathbf{7176.44}$    &$\mathbf{18.18\;(10.757\%)}$ \\
    \cline{2-4}
    & $0.5$  &$\mathbf{7033.25}$    &$\mathbf{9.75\;(5.769\%)}$ \\
    \cline{2-4}
    & $1.0$  &$\mathbf{6776.65}$    &$9.06\;(5.360\%)$ \\
    \cline{2-4}
    & $1.5$  &$\mathbf{7586.59}$    &$\mathbf{7.68\;(4.544\%)}$ \\
    \hline
  \end{tabular}
}
  \caption{Action diversity as a function of $\mu_{\sf line}$ for the hybrid action space $\mcA = \mcA_{\sf line} \bigcup \mcA_{\sf gen}$ with $\mu_{\sf gen}=0$ and $\eta = 0.95$.}
  \label{tab:lineAndGenDiversity}
\end{table}

\begin{table*}[t]
  \centering
  \renewcommand{\arraystretch}{1.2} % Adjust the value as needed
  \scalebox{0.74}{
  {\begin{tabular}{|c|c|c|c|c|c|c|c|}
      \hline
    \thead{Action Space $(|\mcA|)$}  
    & \thead{Agent Type}
    & \thead{Avg. ST} 
    & \thead{$\%$\\ Do-nothing}
    & \thead{$\%$\\ Reconnect}
    & \thead{$\%$\\ Removals}
    & \thead{$\%$\\ Re-dispatch}
    & \thead{Avg. Action\\ Diversity}\\
    \hline
    \hline
    $-$                      &$\noop$ &$4371.91$ &$100$    &$-$     &$-$  &$-$   &$-$ \\
    \hline
    $\mcA_{\sf line}\;(187)$ &$\reco$ &$2813.64$ &$98.73$  &$1.26$    &$-$  &$-$ &$1.235\;(0.66\%)$ \\
    \hline
    \hline
    $\mcA_{\sf line}\;(373)$ &\texttt{milp\_agent}\cite{MILPAGENT:2022}  &$4003.85$ &$15.64$ &$0.88$  &$83.46$  &$-$    &$5.617\;(1.505\%)$ \\
    \hline
    $\mcA_{\sf line} \bigcup \mcA_{\sf gen}\;(-)$ &\texttt{OptimCVXPY}\cite{OptimCVXPY:2022}   &$4976.61$ &$-$ &$-$  &$-$  &$100$    &$8.176\;(-)$  \\
    \hline
    \hline
    \multirow{2}{*}{\rotatebox[origin=c]{0}{\makecell{$\mcA_{\sf line}\;(373)$}}}
        &\multirow{1}{*}{\rotatebox[origin=c]{0}{\makecell{RL\;+ Random Explore}}} 
                      &$4812.88$  &$3.58$ &$20.30$ &$76.08$ &$-$ &$8.323\;(2.231\%)$\\
            \cline{2-8}
        &\multirow{1}{*}{\rotatebox[origin=c]{0}{\makecell{RL\;+ Physics Guided Explore}}} 
                      &$\mathbf{5767.14}$  &$1.86$ &$25.34$ &$72.77$ &$-$ &$\mathbf{16.235\;(4.352\%)}$ \\
    \hline
    \hline  
    \multirow{2}{*}{\rotatebox[origin=c]{0}{\makecell{$\mcA_{\sf line} \bigcup \mcA_{\sf gen}\;(513)$}}}
        &\multirow{1}{*}{\rotatebox[origin=c]{0}{\makecell{RL\;+ Random Explore}}} 
                    &$4880.79$   &$3.13$  &$10.98$  &$14.41$  &$71.45$  &$10.529\;(2.052\%)$ \\
            \cline{2-8}
        &\multirow{1}{*}{\rotatebox[origin=c]{0}{\makecell{RL\;+ Physics Guided Explore}}}
                     &$\mathbf{6095.55}$      &$6.60$   &$14.66$         &$55.92$        &$22.80$  &$\mathbf{15.242\;(2.971\%)}$\\
           \hline
  \end{tabular}}
  }
  \caption{{Performance on the IEEE 118-bus system for various action spaces with $\mu_{\sf line} = \mu_{\sf gen}=0$ and $\eta = 1.0$.}}
  \label{tab:lineAndGen_118}
\end{table*}

%\vspace{-0.3in}

\subsection{{IEEE 118-bus System}}
\label{sec:118}

{
%We next evaluate the performance on a larger system. For a comprehensive evaluation, we performed a random split of Grid2Op scenarios and chose $34$ scenarios for the test set, assigned $450$ to the training set, and set aside a subset for validation.  Similarly, both the architecture and the reward function considered are the same as previously discussed with 
%$\alpha_{n}=9\cdot 10^{-4}$ decayed over $2^{10}$ steps with a mini-batch size of $B=32$, an initial $\epsilon = 0.99$ exponentially decayed to $\epsilon = 0.05$ over $21\cdot 10^{3}$ agent MDP training interaction steps and $\gamma = 0.99$.\\
%For the considered system, following the MDP modeling discussed in Section~\ref{sec: Action Space}, $|\mcA_{\sf line}|=373\;(2L + 1)$. Without loss of generality, we set a uniform line-switch cost $c^{\sf line}_{\ell}=1$ for all lines $\ell \in [L]$. We select a subset consisting of $k=6$ dispatchable generators (out of $G$) with the largest ramp-rates $\Delta G^{\sf max}_{j}$ to design $\mcA_{\sf gen}$ and choose a discretization constant $\delta = 5\; \leq \min \{ \Delta G_{1}^{\sf max}, \dots, \Delta G_{6}^{\sf max} \}$, rendering $|\mcA_{\sf gen}| = 140$.\\
All the results for the IEEE 118-bus system are tabulated in Table~\ref{tab:lineAndGen_118}. Starting from the baselines, we observe that the $\noop$ agent achieves a significantly higher average ST of 4,371 steps, compared to the $\reco$ agent's 2813.64 steps. This observation highlights the importance of strategically selecting~\emph{look-ahead} decisions, particularly in more complex and larger networks. Contrary to common assumptions, the $\reco$ agent's greedy approach of reconnecting lines can instead reduce ST, demonstrating that $\noop$ can be more effective.\\
\indent Focusing on the line switch action space $\mcA_{\sf line}$, we observe that the agent with policy $\pi^{\sf rand}_{\btheta}$ survives 4812.88 steps, a $10.1\%$ increase over baselines, by allocating $76.08\%$ to remedial control actions for line removals. More importantly, our physics-guided policy $\pi^{\sf physics}_{\btheta}$ achieves an average ST of 5767 steps, a $31.9\%$ increase over baselines and a $19.2\%$ improvement compared to $\pi^{\sf rand}_{\btheta}$ with greater action diversity. Fig.~\ref{fig:36bus_interactions} illustrates the number of agent-MDP interactions as a function of training time, showcasing that the physics-guided exploration is more thorough for a given computational budget.\\
\indent Finally, by appropriately designing the generator dispatch action space $\mcA_{\sf gen}$, the $\pi^{\sf physics}_{\btheta}$ policy achieves the highest survival time of 6,095.55 steps, outperforming all baselines, by allocating $22.8\%$ of actions to generator dispatch. This demonstrates the benefits of combining line-switching and generator adjustments by leveraging the proposed generator action space modeling technique. All the observations corroborate those observed for the Grid2Op 36-bus system.
}

% \begin{figure}[t]
% \centering
% \includegraphics[width=0.75\linewidth]{Cascading Failure Mitigation/Figures/118bus_EnvInteractions.pdf}
% \caption{{Agent$-$MDP interactions for the IEEE 118-bus system with $\mcA = \mcA_{\sf line}$ and $\mu_{\sf line}=\mu_{\sf gen}=0$ for $\eta = 1.0$.}}
% \label{fig:118bus_interactions}  
% \end{figure}

{\subsection{Discussions}}
\label{sec: Discussions}

\subsubsection{{Scalability}}
{We enhance network resilience by primarily altering the network topology as a preventive measure to alleviate overflows, which necessitates the re-computation of the LODFs, which has a time complexity $\mcO(N^{3})$. It is noteworthy that this re-computation arises only under critical states, specifically when $\rho_{\ell_{\sf max}}[n] \geq \eta$, which occurs infrequently. Additionally, these re-computations are confined to the agent training phase. Once the neural network weights $\btheta$ have been trained to parameterize the policy $\pi_{\btheta}$, sensitivity factor computations are not needed at inference time.
}

\subsubsection{{Choice of RL Algorithm}}
{One takeaway of this paper is to showcase how physical signals and domain knowledge can be integrated into RL frameworks. This establishes the benefits of integrating domain knowledge into $\epsilon$-greedy based RL algorithms for a more directed exploration in complex environments. The same idea of creating a physics-guided exploration is broadly applicable to other advanced RL algorithms, such as policy gradients. Rather than randomly initializing the policy network's weights $\btheta$, pre-training the network with the ultimate objective of maximizing the action probabilities associated with the effective action set $\mcR^{\sf eff}_{\sf line}[n] \subseteq \mcA$ will render the agent sensitivity factor-aware during exploration. It is noteworthy that the broader applicability of this framework to other RL algorithms is also underscored by the recent study in~\cite{meppelink:2023}, demonstrating improved convergence by incorporating hop-based electrical distance into MCTS.
}

\subsubsection{{Potential Applications}}
{Beyond the practical benefits, our framework can serve as a basis for designing exploration policies suited for other power system control objectives:}\\
{\textbf{Efficient Action Space Pruning:} Existing approaches to managing expansive action spaces (e.g., bus-splitting) rely on exhaustive search methods, where~\emph{each} action is~\emph{simulated} from each state in the training data scenarios. Subsequently, the most frequently chosen actions are then selected for pruning~\cite{wang2016dueling, AAAI:2023, dorfer:2022, marot:2021}. This is computationally intensive. 
The recent advancements such as bus-split factors~\cite{BSDF:2024}, show promise in streamlining this process. Specifically, our framework facilities the design of an efficient action set $\mcR^{\sf eff}[n] \subseteq \mcA$ that requires implementing only these selected actions and enhances the computational efficiency of action space pruning.}\\
{\textbf{Simultaneous Action Sequences:} Exploring simultaneous switching of lines and generator adjustments holds great promise for optimizing power system operations. However, designing such sequences of simultaneous actions is challenging due to the combinatorially many choices in the action set $\mcA_{\sf line} \times \mcA_{\sf gen}$ at each time instant $n$. Leveraging the~\emph{linearity} of sensitivity factors, our framework offers an efficient approach to exploring effective combinations of remedial actions.}

%\vspace{-0.1in}

\section{Conclusion}
\label{sec:Conclusion}

In this paper, we have introduced a physics-guided RL framework to determine sequences of effective real-time remedial control decisions for blackout mitigation. The approach, focused on transmission line-switches and generator adjustments, utilizes linear sensitivity factors to enhance RL exploration during agent training. Simulations on the open-source Grid2Op platform demonstrate that strategically removing transmission lines, alongside multiple real-time generator adjustments, provides a more effective long-term remedial control strategy. Comparative analyses on the Grid2Op 36-bus and~{the IEEE 118-bus networks} highlight the superior performance of our framework against relevant baselines. %Future work involves leveraging sensitivity factors for bus-split action space design and graph neural networks with time-varying filter functions to learn a remedial control policy.

\section{Acknowledgements} 
A. Dwivedi would like to thank N. Virani for the useful discussions in the early stages of this work.

\bibliographystyle{IEEEtran}
\bibliography{CFM}

\end{document}